\documentclass[aps,prx,twocolumn,showpacs,floatfix,notitlepage]{revtex4-1}
\usepackage[latin9]{inputenc}
\usepackage{color}
\usepackage{verbatim}
\usepackage{units}
\usepackage{amsmath}
\usepackage{amssymb}
\usepackage{graphicx}
\usepackage{graphicx}  
\usepackage{epstopdf}
\usepackage{bm}        
\usepackage{amsmath}
\usepackage{amssymb}   
\usepackage{array}
\usepackage[caption=false]{subfig}
\usepackage[bookmarks=false,linkcolor=blue,urlcolor=blue,colorlinks,citecolor=blue]{hyperref}
\usepackage{dsfont}
\usepackage{appendix}

\makeatletter
\@ifundefined{textcolor}{}
{%
 \definecolor{BLACK}{gray}{0}
 \definecolor{WHITE}{gray}{1}
 \definecolor{RED}{rgb}{1,0,0}
 \definecolor{GREEN}{rgb}{0,1,0}
 \definecolor{BLUE}{rgb}{0,0,1}
 \definecolor{CYAN}{cmyk}{1,0,0,0}
 \definecolor{MAGENTA}{cmyk}{0,1,0,0}
 \definecolor{YELLOW}{cmyk}{0,0,1,0}
}

\usepackage{dcolumn}
\usepackage{bm}
\usepackage{setspace}
\usepackage{fancybox}
\usepackage{listings}
\usepackage{url}
\usepackage{pdfsync}
\usepackage{float}

\usepackage{subfig}

\begin{document}

\title{Imprint of topological degeneracy in quasi-one-dimensional fractional quantum Hall states}

\author{Eran Sagi}
\affiliation{Department of Condensed Matter Physics, Weizmann Institute of Science,
Rehovot, Israel 76100}
\author{Yuval Oreg}
\affiliation{Department of Condensed Matter Physics, Weizmann Institute of Science,
Rehovot, Israel 76100}
\author{Ady Stern}
\affiliation{Department of Condensed Matter Physics, Weizmann Institute of Science,
Rehovot, Israel 76100}
\author{Bertrand I. Halperin}
\affiliation{Department of Physics, Harvard University, Cambridge MA 02138, USA}

\begin{abstract}
We consider an annular superconductor-insulator-superconductor Josephson-junction,
with the insulator being a double layer of electron and holes at Abelian fractional quantum Hall states of identical fillings.
When the two superconductors gap out the edge modes, the system has a topological ground state degeneracy in the thermodynamic limit akin to the fractional quantum Hall degeneracy on a torus. In the quasi-one-dimensional limit, where the width of the insulator becomes small, the ground state energies are split. We discuss several implications of the topological degeneracy that survive the crossover to the quasi-one-dimensional limit. In particular, the Josephson effect shows a $2\pi d$-periodicity, where $d$ is the ground state degeneracy in the 2 dimensional limit. We find that at special values of the relative phase between the two superconductors there are protected crossing points in which the degeneracy is not completely lifted. These features occur also if the insulator is a time-reversal-invariant fractional topological insulator. We describe the latter using a construction based on coupled wires.
Furthermore, when the superconductors are replaced by systems with an appropriate magnetic order that gap the edges via a spin-flipping backscattering, the Josephson effect is replaced by a spin Josephson effect.

\end{abstract}
\pacs{73.43.-f,73.21.Hb,03.65.Vf, 74.78.Fk} 
\maketitle

\section{Introduction}

\label{sec:introduction}

One of the hallmarks of the fractional quantum Hall effect (FQHE) is that
if the two-dimensional electron system resides on a manifold with a
nontrivial topology, it will have a ground state degeneracy which depends
on the topology
\cite{Wen1990}. For a fractional
quantum Hall state on an infinite torus, the degeneracy of the ground
state equals the number of topologically distinct fractionalized quasi-particles
allowed in that state. Since this degeneracy is topological, it does
not originate from any symmetry, and in particular does not require
the absence of disorder. Furthermore, no local measurement may distinguish
between the degenerate ground states.

When the torus is of large but finite size, the degeneracy is split,
but the splitting is exponentially small in $L,$ where $L=\min\left\{ L_{x},L_{y}\right\} $
and $L_{x},L_{y}$ are the two circumferences of the torus. In the
thin torus regime, where one circumference of the torus is infinite
and the other is smaller or comparable to the magnetic length, the fractional
quantum Hall state crosses over into a charge density wave (CDW),
and the degenerate ground states correspond to different possible
phases of the CDW \cite{Bergholtz2005, Bergholtz2006, Seidel2005}. In that regime a local impurity may pin the charge density wave and lift the degeneracy between the ground states. Equivalently, a local  measurement is able to identify
the phase of the CDW, and hence the ground state.

\begin{figure*}
\begin{center}
\subfloat[\label{fig:annulus_uniform}]{\includegraphics[scale=0.6]{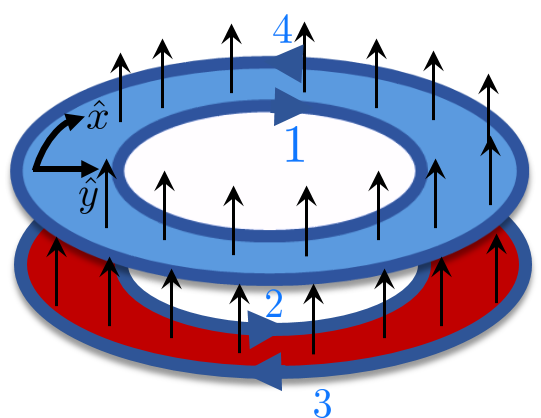}}
\subfloat[\label{fig:system}]{\includegraphics[scale=0.4]{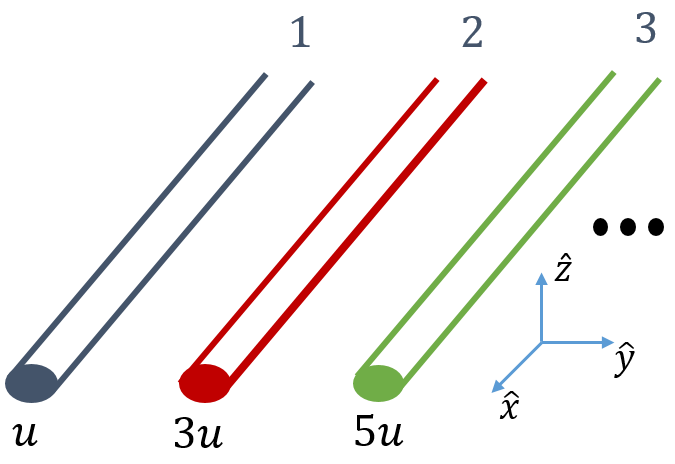}}
\subfloat[\label{fig:relative angle}]{\includegraphics[scale=0.5]{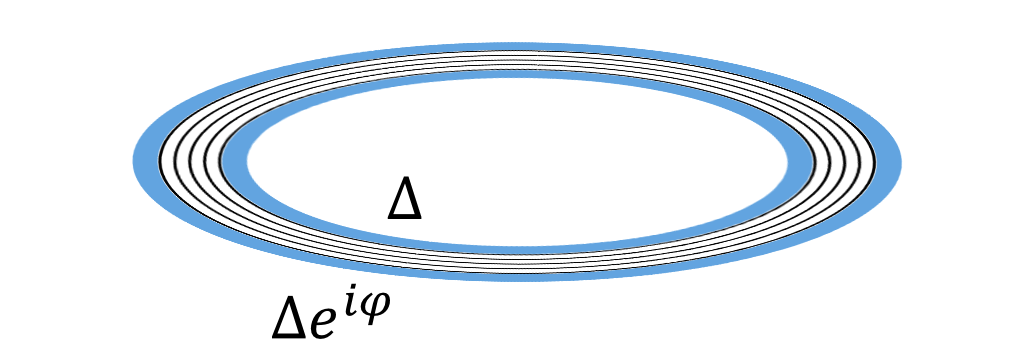}}

\end{center}

\protect\protect
\caption{(a) The first realization we consider is that of an electron annulus (blue) and a hole annulus (red) under the action of a uniform magnetic field. It is evident that coupling the annuli's edges forms the topology of a torus. The second realization we suggest is that of a fractional topological insulator. Fig. (b) shows a possible model for a fractional topological insulator. We have an array of $N$ wires, with a strong spin-orbit coupling. The spin orbit coupling is linear with the wire index $n$.  The similarity of the resulting spectrum
(see Fig. (\ref{fig:spectrum_no_tunneling}) below) to the one corresponding to
the wires construction of quantum Hall states suggests an equivalence
to two quantum Hall annuli subjected to opposite magnetic fields (each
annulus corresponds to a specific spin). The use of the wires construction enables us to include interaction effects using a bosonized Tomonaga-Luttinger liquid theory for the description of the wires.
(c) The edge modes of the two above models can be gapped out by proximity coupling to superconductors. In the case of a thin (quasi-1D) system, the phase difference between the inner and the outer superconductors leads
to a Josephson effect mediated by tunneling across the region of a fractional quantum Hall double layer
 or a fractional topological insulator.
The spectrum as a function
of the phase difference $\varphi$ is depicted in Fig. (\ref{fig:spectrum with flux}) below. The edge modes can also be gapped using proximity to magnets, in which case one can measure the spin-Josephson effect.} 
\end{figure*}

In this work we consider two systems that are topologically equivalent to a torus, and - unlike the torus - are within experimental reach.
The first is that of an annular shaped electron-hole double-layer in which
the electron and hole densities are equal, and are both tuned to the
same FQHE state (see Fig. (\ref{fig:annulus_uniform})). In the absence
of any coupling between the layers, both the interior edge and the
exterior edge of the annulus carry pairs of counter-propagating edge
modes of the electrons and the holes. These pairs may be gapped by
means of inter-layer back-scattering, resulting in a fully gapped
system with the effective topology of the torus. In fact, this system
is richer than a seamless torus, since the interior and exterior edges
may be gapped in different ways. In particular, gapping the counter-propagating
edge modes by coupling them to a superconductor
 may have interesting consequences. Some of these consequences are central to the current paper.

The second realization we consider is that of a two dimensional time-reversal-invariant fractional topological
insulator \cite{Levin2009}. To be concrete, we assume that it is constructed of wires subjected to spin-orbit coupling and electron-electron
interaction (see Fig. (\ref{fig:system})). In this realization, electrons
of spin-up form a FQHE state of filling factor $\nu$, and electrons
of spin-down form a FQHE of filling factor $-\nu$. Similar to the particle-hole case, the edges
carry pairs of counter-propagating edge modes with opposite spins that may be gapped in
different ways. Remarkably, when the edge modes are gapped by being
coupled to superconductors, the system is invariant under time-reversal,
yet topologically equivalent to a FQHE torus.

We use these realizations of a toroidal geometry and
their inter-relations to investigate the transition of a fractional
quantum Hall system from the thermodynamic two-dimensional to the
quasi-one dimensional regime of a few wires. In particular, we find signatures of
the topological ground state degeneracy of the two-dimensional (2D) limit (akin to that of fractional quantum Hall states on a torus) that survive the transition
to the quasi one-dimensional (1D) regime and propose experiments in which these signatures may
be probed. For example, for an Abelian fractional quantum Hall state, we find a $2\pi d$-periodic Josephson effect, where $d$ is the degeneracy in the 2D thermodynamic limit. We note that related ideas were explored in Ref. \cite{Iadecola2014}, where it was suggested that a signature of the ground state degeneracy can be found by measuring the heat capacity.\\
\phantom{---}The structure of the paper is as follows: in Sec.~\ref{sec:Summary} we summarize the physical ideas and the main results of the paper. In Sec. \ref{sec:model} we define the systems in more detail and identify the topological degeneracy in the thermodynamic limit. In Sec. \ref{sec:crossover}, we discuss the quasi one-dimensional regime, and point out observable signatures of the topological degeneracy in that regime. Our discussions in these sections focus on the $\nu=1/3$ case. In Sec. \ref{sec:extension} we discuss how the results of the previous sections are generalized to other Abelian QHE states.

\section{The main results and the physical picture}
\label{sec:Summary}

\subsection{The systems considered}
\label{sub:system}

The electron-hole double-layer system is conceptually simple to visualize
(see Fig.~(\ref{fig:annulus_uniform})). We consider an electron-hole
double-layer shaped as an annulus with equal densities of electrons and
holes, and a magnetic field that forms FQHE states of $\pm\nu$ in
the two layers. The system breaks time reversal symmetry, but its
low energy physics satisfies a particle-hole symmetry. For most of
our discussion we focus on the case $\nu=1/3$. In that case each
edge carries a pair of counter-propagating $\nu=1/3$ edge modes.
The edge modes may be gapped by means of normal back-scattering (possibly
involving spin-flip, induced by a magnet) or by means of coupling
to a superconductor.
  In line with common notation, we refer to these two ways as $F$ and $S$ respectively.

To model the fractional topological insulator we consider an array of $N$ coupled quantum wires of length
$L_{x}$, each satisfying periodic boundary conditions (Fig. (\ref{fig:system})).
The wires are subjected to a Rashba spin-orbit coupling, and we consider
a case in which the spin-orbit coupling constant in the $n$'th wire
is proportional to $2n-1$ (similar to the model considered by Ref. \cite{Klinovaja2014}). Effectively, this form of spin-orbit coupling
subjects electrons of opposite spins to opposite magnetic fields.
While this particular coupled-wire model of a time reversal invariant topological insulator does not naturally allow for the regime of a large $N$, other realizations, such as those  proposed in Ref. \cite{Sagi2014a,Klinovaja2014}, allow for such a regime. These realizations require more wires in a unit cell, and are therefore more complicated that the one considered here. Most of the results of our analysis are independent of the specific realization of the fractional topological insulator, and we present the analysis for the realization that is simplest to consider.

For non-interacting electrons, the spectrum of the array we consider takes the form shown in Fig. (\ref{fig:spectrum_no_tunneling}).
Single-electron tunneling processes (which conserve spin) gap out the spectrum in all but
the first and last wires, which carry helical modes (Fig. (\ref{fig:spectrum_no_tunneling-1})).
If the chemical potential is tuned to this gap, then in the limit
of large $N$ the system is a topological insulator (TI), and therefore
the gapless edge modes are protected by time-reversal symmetry and
charge conservation \cite{Kane2005}. This construction is then equivalent to two electron QH
annuli with opposite magnetic fields.

The edge modes may be gapped by coupling the two external wires ($n=1$
and $n=N)$ to a superconductor or to a system with appropriate magnetic order. A Zeeman field that is not collinear with the spin-orbit coupling direction is necessary to couple the different spin directions. Moreover, in our coupled-wires model the spin-up and the spin-down electrons at the $n=N$ edge have different Fermi-momenta, so that edge would not be gapped by a simple ferromagnet. In order to conserve momentum one would need to introduce a periodic potential that could modulate the coupling to the ferromagnet at the appropriate wave vector, or one would need to use a spiral magnet with the appropriate pitch. In more sophisticated wire models, such as those discussed by Refs. \cite{Sagi2014a,Klinovaja2014}, or in actual realizations of topological insulators, the two edge modes can have the same Fermi momenta, so a simple ferromagnet can be used.

In order to construct a fractional topological insulator, we first
tune the chemical potential such that the density is reduced by a
factor of three, to $\nu=1/3$. For an array of wires in a magnetic
field and spinless electrons, Kane et al. \cite{Kane2002} have introduced an interaction
that leads to a ground state of a FQHE $\nu=1/3.$ Furthermore, they
argued that there is a range of interactions that will flow to the
topological phase described by this state \cite{Kane2002,Teo2011,Oreg2013}.
Here we show that the same interaction, if operative between electrons
of the same spin only, leads to a formation of a fractional topological
insulator, i.e., to the spin-up electrons forming a $\nu=1/3$ state
and the spin-down electrons forming a $\nu=-1/3$ state. Note that the same type of interaction terms were used by several authors to construct various 2D fractional topological states \cite{Neupert2014,Sagi2014a,Klinovaja2014}, and 1D fractional states \cite{Oreg2013,Klinovaja2013b,Klinovaja2013,Orth2014}.

Our analysis is based on bosonization of the wires' degrees of freedom,
and a transformation to a set of composite chiral fields, that may
be interpreted as describing fermions at filling $\nu=1$. In terms
of the composite fields, one can repeat the process which led to a
gapping of the non-interacting case either by normal or by superconducting
mechanisms. In terms of the original electrons, these mechanisms involve
multi-electron processes, which either conserve the number of electrons
or change it by a Cooper pair.

Both the electron-hole double-layer and spin-orbit wire system have counter propagating edge modes. They are distinct, however, in a few  technical details. An electron-hole double layer system has been realized before in several materials, such as GaAs quantum wells and graphene. The requirements we have here - no bulk tunneling, sample quality that is sufficient for the observation of the fractional quantum Hall effect, and a good coupling to a superconductor or a magnet - are not easy to realize, but are not far from experimental reach \cite{Gorbachev2012,Kou2014,Eisenstein2014}. In addition, we assume that the two layers are far enough such that inter-layer interactions do not play an important role, but close compared to the superconducting coherence length to enable pairing on the edges.

The array of wires we describe can in principle be formed using semi-conducting wires such as InAs and InSb \cite{Mourik2012,Das2012,Deng2012}, where variable Rashba spin-orbit coupling could be achieved by applying different voltages to gates above the wires. We stress that the wires construction is nothing but a specific example of a fractional topological insulator, and that any fractional topological insulator is expected to present the effects we discuss. Two-dimensional topological insulators were conclusively observed \cite{Bernevig2006a,Konig2007,Liu2008,Roth2009,Nowack2013,Du2013,Spanton2014}, and more recently proximity effects to a superconductor were demonstrated on their edges \cite{Knez2012,Hart2013,Pribiag2014}. However, fractionalization effects due to strong electron-electron interaction were not observed yet in these systems and are less founded theoretically.

We emphasize that our construction, which is equivalent to a single layer quantum Hall state on a torus, is different  from toroidal geometry of a double layer quantum Hall state.

\subsection{Ground state degeneracy and its fate in the transition to one dimension}

In Sec.~\ref{sec:model} we investigate the topological degeneracy
of the ground state in the 2D thermodynamic limit. Using general
arguments, we find that the degeneracy depends on the gapping mechanism
of the edges: when both edges are gapped by the same mechanism, be
it proximity coupling to a superconductor or to a magnet, the
topological degeneracy is three, as expected. However, if one edge
is gapped using a superconductor and the other is gapped using a magnet
the ground state of the system is not degenerate.

Physically, the degeneracy is most simply understood in terms of the
charge on the edge modes. For an annular geometry there are two edges,
in the interior and the exterior of the annulus, and therefore four
edge modes with four charges, $q_{1},q_{2},q_{3},$ and $q_{4}$ (here we
use the subscript 1,2 to denote the two counter-propagating edge modes
on the interior edge, and 3,4 to denote the modes on the exterior
edge. Edges 1 and 4 belong to one layer (or one spin direction) and edges 2 and 3
belong to the other layer (other spin direction); see Fig. (\ref{fig:annulus_uniform})).  It will be useful below to distinguish between the integer part of $q_i$, which we denote by $n_i$, and the fractional part denoted by $f_i$, to which we assign the values $f_i=-1/3,0,1/3$, such that $q_i=n_i+f_i$.

When a pair of counter-propagating edge modes, say with charges $q_{1},q_{2}$, is gapped by normal back-scattering
of single electrons, their total charge $q_{1}+q_{2}$ is conserved.
Since there is an energy cost associated with the total charge, it
assumes a fixed value for all ground states.
(The tunneling between the edges gaps the system and makes it incompressible, leading to an energy cost associated with a change of the total charge.)
For simplicity, we
fix this value to be zero, making $q_{1}=-q_{2}$. A strong back-scattering term
makes  $n_1-n_2$ strongly fluctuating {\it but leaves the fractional part $f_1=-f_2$ fixed}.  As a consequence, there are three topological sectors of states that are not coupled by electron tunneling, characterized by $f_1$ being
0, 1/3 or -1/3.

Since each of the layers (in the double-layer system) or each spin direction
(in the spin-orbit-coupling system) must have an integer number of electrons, the
sums $q_{1}+q_{4}$ and $q_{2}+q_{3}$ must both be integers. This
condition couples the fractional parts of the charges on all edges.
Combining all constraints, we find that when both edges are gapped
by a normal backscattering, the following conditions should be fulfilled
 \begin{equation}
 f_1=-f_2, \;\; \; f_3=-f_4, \label{eq:const1}
 \end{equation}
  \begin{equation}
 \label{eq:const2}
  f_1=-f_4, \;\;\; f_2=-f_3. 
 \end{equation}
 There are three solutions for these equations describing three ground states, with  $f_l=\left(-1\right)^{l}\frac{p}{3}$, where $p$ may take the values $0,1,-1$ and $l=1,2,3,$ and $4$.
 When both edges are gapped by a
superconductor, $f_2$ and $f_4$ change sign in Eq. (\ref{eq:const1}) and  the fractional parts satisfy $f_{1}=f_{2}=-f_{3}=-f_{4}=p/3$, Finally, when one
edge is gapped by a superconductor and the other by normal back-scattering,
only one of the two equations labeled ~(\ref{eq:const1}) change sign and the only possible solution is $f_l=0$ so that all $q$'s must be integers, and the ground state is unique.

Formally, the
degeneracy of the ground state may be shown by an explicit construction of two unitary operators,
$U_{x}$ and $U_{y}$, that commute with the low-energy effective Hamiltonian and satisfy
the operator relation
\begin{equation}
U_{x}U_{y}=U_{y}U_{x}e^{\frac{2\pi}{3}i}.\label{eq:O and U commutation}
\end{equation}
The existence of a matrix representation of this relation, acting
within the ground state manifold, requires a degenerate subspace of
minimal dimension 3.

We construct such operators for the electron-hole system under the assumption that the only active degrees of freedom are those of the edge, and for the coupled wire system when we confine ourselves to an effective Hamiltonian. For both cases, one of these operators, say $U_{x},$ measures the $f_l$'s and the other operator, $U_{y},$ changes
the $f_l$'s by  $\pm\nicefrac{1}{3}$ (the sign depends on $l$ and on the
type of gapping mechanism). We choose to work with a representation of $U_x,U_y$ in which  both operators, projected to subspace of ground states, are independent of position.

Even when $L_x$ is infinite, a finite $L_y$ splits the degeneracy. The source of lifting of the degeneracy is tunneling of quasi-particles between
the two edges of the annulus, i.e., tunneling of quasi-particles from
the first to the last wire. More precisely, we find that as long as
the bulk gap does not close, the only term that may be added to the
low-energy Hamiltonian is of the form
\begin{equation}
\lambda U_y+\lambda^* U_y^\dagger
\label{deviation}
\end{equation}
This term is generated by high orders of perturbation theory that lead to a transfer of quasi-particles between edges.
The amplitude $\lambda$ decays exponentially with the width of the system. For the wires realization this translates to an exponential decay with $N$, the number of wires.
Other factors that determine the magnitude and phase of $\lambda$ are elaborated on in the next subsection.

If $L_x$ is also finite, there will be additional terms in the Hamiltonian proportional to $U_x$ and $U_x^\dagger$, with coefficients that fall off exponentially in $L_x$. The physical explanation of these terms is that when $L_x$ is finite,  root-mean-square fluctuations in the total charge in an edge mode are not infinite, but are proportional to $L_x^{1/2}$. This leads to energy differences between states with different values of the fractional charge $f_l$ that decrease exponentially with increasing $L_x$

\subsection{Remnants of the degeneracy in the quasi-one dimensional regime}

\label{sub:proposed experiment}

The topological degeneracy is lifted in the transition from a two-dimensional system to a quasi-one dimensional one, but it leaves behind an imprint which can in principle
be measured. This is seen when we add another parameter to the Hamiltonian.
For a torus, this parameter may be the flux within the torus. For
the systems we consider here, when gapped by one superconductor
at the interior edge and one superconductor at the exterior edge,
this parameter may be the phase difference $\varphi$ between the
two superconductors. In this case the fractional quantum Hall torus
forms the insulator in a superconductor-insulator-superconductor
Josephson junction.

\begin{figure}
\includegraphics[scale=0.24]{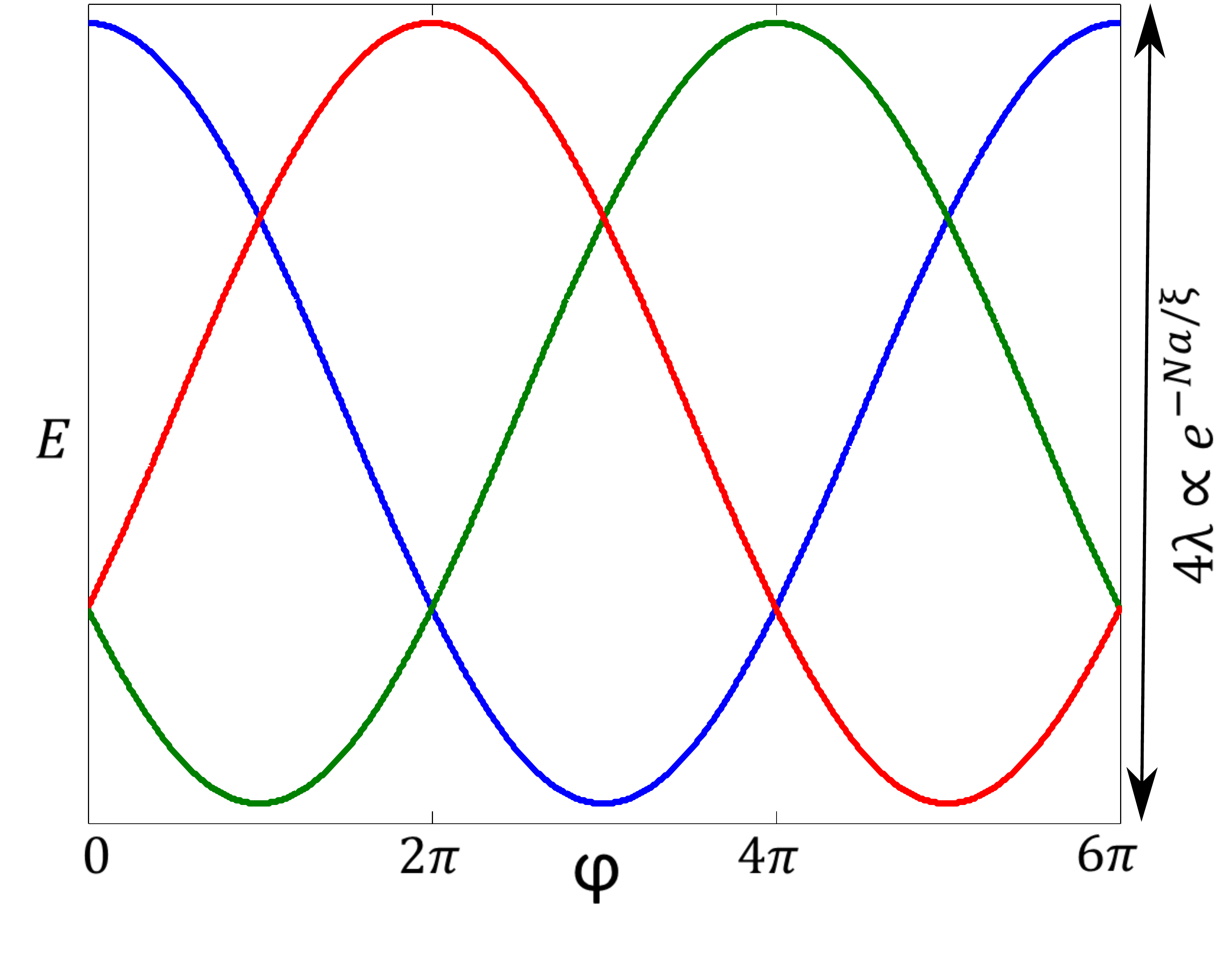}\protect\protect\caption{\label{fig:spectrum with flux}The spectrum of the three low energy
states as a function of the phase difference $\varphi$ between the two superconductors (see text for elaboration). The amplitude of oscillations
falls exponentially with the number of wires $N$. For a finite
$N$, each eigenstate has a periodicity of $6\pi$. At the special points $\varphi=\pi n$ the spectrum remains 2-fold degenerate. If the system is of finite length $L_x$, the degeneracy at these points is lifted by a term that is exponentially small in $L_x$.}
\end{figure}

{The dependence of the spectrum on these parameters is encoded in the amplitude $\lambda$ of Eq. (\ref{deviation}). In particular, since the tunneling charge is $2/3$ of an electron charge, which is $1/3$ of a Cooper pair, we find that the tunneling amplitude at the point $x$ along the junction is proportional to the phase factor $e^{i\varphi(x)/3}$, where $\varphi(x)$ is the phase difference between the two superconductors at the point $x$. For the fractional topological insulator, no magnetic flux is enclosed between the superconductors, and the equilibrium phase difference does not depend on $x$. In contrast, for the electron-hole quantum Hall realization the magnetic flux threading the electron-hole double layer makes $\varphi(x)$ vary linearly with $x$, such that the phase of the tunneling amplitude winds as a function of the position of the tunneling.
The amplitude $\lambda$ of Eq. (\ref{deviation}) is an integral of contributions from all points at which the superconductors are tunnel-coupled,
\begin{equation}
\lambda=\int dx T(x)
\label{localtunnelingintegrated}
\end{equation}
where $T(x)$ is the local tunneling amplitude.

When the superconductors are tunnel-coupled only at a single point (say $x=0$), such that $T (x)\propto \delta (x)$, the spectrum of the three ground states as a function of $\varphi$, which is now the argument of $T(x=0)$ can be written in the explicit form
 \begin{equation}
\Delta E_{\alpha} = 2t_{0}\cos\left(\frac{\varphi-2\pi \alpha}{3}\right), \label{eq:energies of ground states}
\end{equation}
where $\alpha=0,1,-1$ enumerates the ground states.  This is shown in Fig. (\ref{fig:spectrum with flux}).  }

While the amplitude $t_0$ is exponentially small in the width $L_y$, or in the number of wires $N$, we find that the spectrum as a function of the
phase difference across the junction has points of avoided crossing
in which the scale of the splitting between the two crossing states
is proportional to $e^{-L_{x}/\xi_x}$, i.e., is exponentially small
in $L_{x}$ (here $\xi_x$ is a characteristic
scale which depends on the microscopics). Thus, in the quasi-one-dimensional
regime, where $L_{y}$ or $N$ are small but $L_{x}$ is infinite,
the three states are split, but cross one another at particular values
of $\varphi$.

Remarkably, this crossing cannot be lifted by any perturbation
that does not close the gap between the three degeneracy-split ground
states and the rest of the spectrum. This lack of coupling between
these states result from the macroscopically different Josephson current (from the inner edge to the outer edge)
that they carry.
The Josephson
junction formed between the two superconductors will show a $6\pi$-
periodic DC Josephson effect for as long as the time variation of
the phase is slow compared to the bulk energy gap, but fast compared to
a time scale that grows as $e^{L_{x}/\xi_x}$. This Josephson current distinguishes between the three ground states. This current oscillates as a function of the position of the tunneling point for an electron-hole quantum Hall system and is position-independent for the fractional topological insulator.

When tunneling between edges takes place in more than one point, $T(x)$ in (\ref{localtunnelingintegrated}) is non-zero at all these points, and has to be integrated. A particularly interesting case is that of a uniform junction. In that case $T(x)$ and the Josephson current are constant for the fractional topological insulator, while in the electron-hole double-layer the phase of $T(x)$ winds an integer number of times due to the magnetic flux between the superconductors, and the Josephson current averages to zero.

A magnetic coupling between the electron and hole layers, or between electrons of the two spin directions  may lead to a ``(fractional) spin Josephson effect", in which spin current takes the place of charge current in the Josephson effect \cite{Nilsson2008, Jiang2013, Pientka2013}.
In this case, assuming that the spin up and down electrons are polarized in the $z$ direction, coupling between the edge modes occurs by a magnet that exerts a Zeeman field in the $x-y$ plane. The role of the phase difference in the superconducting case is played here by the relative angle between the magnetization at the interior and exterior edge, but an interesting switch between the two systems we consider takes place. In the electron-hole quantum Hall case
the direction of the magnetization is uniform along the edges and a uniform and opposite electric current flows in the two layers.

For the fractional topological insulators the edges are gapped only when for one of the edges the direction of the magnetization in the $x-y$ plane winds as a function of position. As a consequence, in our coupled-wire model the spin current oscillates an integer number of oscillations along the junction, and thus averages to zero.

Our discussion may be extended beyond the case of $\nu=1/3$. For
Abelian states, we find that the periodicity of the Josephshon effect
is $2\pi/e^{*},$ where $e^{*}$ is the smallest fractional charge
allowed in the state. In any Abelian state, this is also $2\pi$ times the degeneracy of
the ground state in the thermodynamical limit.

\section{Ground state degeneracy in the thermodynamic 2D limit \label{sec:model}}

In this section we derive in detail the degeneracy of the ground state in the
thermodynamic two-dimensional limit of the two systems we consider.

\subsection{Description in term of edge modes only}
\label{sec:edge modes only}
The systems we consider have two edges, each of which carrying a pair of counter-propagating edge modes.

In
the absence of coupling between the layers, the bosonic Hamiltonian
of the edges is composed of the kinetic term

\begin{equation}
H_{0}=\frac{v}{2}\int dx  \sum_{l=1,2,3,4} \left(\partial_{x}\chi_{l}\right)^{2}.\label{eq:kinetic part}
\end{equation}

Here we assumed all edge velocities to be the same and neglected small-momentum
interaction between the edges, for simplicity.

The fields $\chi_{i}$ satisfy the commutation relation
\begin{equation}
\left[\chi_{l}(x_l),\chi_{j}(x_j)\right]=i\frac{1}{3}(-1)^{l}\pi\delta_{lj} {\text{sign}} (x_l-x_j)+i\frac{1}{9}\pi {\text{sign}} (l-j).\label{eq:commutation of phi-1}
\end{equation}

Coupling between the edge modes has the form
\begin{equation}
H_{1}=\lambda\int dx\cos3\left(\chi_{l}\pm\chi_{j}\right),\label{eq:cosine perturbation}
\end{equation}
where $l,j=1,2$ for the interior edge and $l,j=3,4$ for the exterior
edge. The plus sign refers to superconducting coupling and the minus
sign to normal back-scattering. The edge is gapped when the coupling
constant $\lambda$ is large, which we assume to be the case.

The charge on the $l$'th edge modes is related to the winding of $\chi_{l}$,
namely $q_{l}=(-1)^l\frac{1}{2\pi}\int dx\partial_{x}\chi_{l}(x)$, where $q_{l}$
is the charge in units of the electron charge. For uncoupled edge
modes, the charges $q_{l}$ are quantized in units of the quasi-particle
charge, 1/3. When two edge-modes are coupled through a normal or superconducting coupling, the charge on each edge heavily fluctuates. However, due
to the fact that only whole electrons may be transferred between edge
modes on different layers, or between edge modes and an adjacent superconductor,
the operators $e^{i 2\pi q_{l}}$ commute with both parts of the Hamiltonian
Eqs. (\ref{eq:kinetic part}) and (\ref{eq:cosine perturbation}). We
therefore characterize the different states according to these operators,
{\it i.e.}, according to the fractional part of the charge on the various
edges. The fact that the total charge on each layer is
an integer gives the two general constraints
\begin{align}
\exp\left[i2\pi\left(q_{1}+q_{4}\right)\right] & =1,\nonumber \\
\exp\left[i2\pi\left(q_{2}+q_{3}\right)\right] & =1,\label{eq:first two equations-1}
\end{align}
regardless of the mechanism for coupling the edges. Two other relations come from
energy considerations, which depend on the gapping mechanism. For
the case where the two edges are gapped using a superconductor it is energetically favorable to
form singlets, such that
\begin{align}
q_{1} & =q_{2},\nonumber \\
q_{3} & =q_{4}.\label{eq:equations-SC-1}
\end{align}
Notice that if Eq. (\ref{eq:equations-SC-1}) is not satisfied, the edge carries a non-zero spin which cannot be screened by the superconductor. This configuration is therefore energetically costly.

In the case where both edges are gapped by normal back-scattering
processes, which we refer to as the FF
case, it is energetically favorable to preserve
total charge neutrality because an insulating  magnet cannot screen charge.
This gives us the conditions
\begin{align}
q_{1} & =-q_{2},\nonumber \\
q_{3} & =-q_{4}.\label{eq:equations-FM-1}
\end{align}

Altogether, then, for the SS and FF gapping mechanisms, there are three possible
values for $e^{i2\pi q_{1}}$, namely $1,\, e^{i2\pi/3},e^{i4\pi/3},$
and the eigenvalue of this operator fixes the values of all operators
$e^{i2\pi q_{l}}$ (for $l=2,3,4)$. These operators are of course equal to the $e^{i2\pi f_l}$ introduced above. In fact, the operators $e^{i 2\pi q_{l}}$
may all serve as the unitary operators $U_{x}$ from Eq. (\ref{eq:O and U commutation}).
To establish a ground state degeneracy, we need to find an operator
that commutes with the Hamiltonian and varies $U_x$.
This operator is the one that transfers a charge of 1/3 in each
layer (for the SS case), or charges of $\frac{1}{3},-\frac{1}{3}$
(for the FF case) from the interior to the exterior. For example, if we choose $U_x=e^{2\pi i q_1}$ then,
\begin{equation}
U_{y}=\exp\left[-i\left(\chi_{1}\pm\chi_{2}-\chi_{3}\mp\chi_{4}\right)\right].\label{eq:Uy in terms of chi}
\end{equation}
Here the upper sign refers to superconducting coupling and the lower sign to coupling to a magnet. The fields $\chi_i$ in (\ref{eq:Uy in terms of chi}) are all to be evaluated at the same point $x$.

It is easy to see that this assignment of $U_{x},U_{y}$ satisfies
Eq. (\ref{eq:O and U commutation}), thus establishing the ground
state degeneracy of the Hamiltonian in Eqs. (\ref{eq:kinetic part}) and (\ref{eq:cosine perturbation})
for the cases of SS and FF gapping mechanisms. In the case where the two edges
are gapped using different mechanisms (FS or SF),
the only solution is the one where $e^{i2\pi q_{l}}$=1
(for $l=1,2,3,4)$, and the ground state is therefore non-degenerate.

For a finite system the three-fold degeneracy is split. In particular, in the quasi-1D regime in which $L_x$ is infinite and $L_y$ is finite, the splitting is a consequence of tunnel coupling between the interior and the exterior. This regime will be explored below.

Before doing that, however, we introduce the coupled wires
system and study its ground state degeneracy directly.

\subsection{The coupled wires construction for a Fractional Topological Insulator}
\label{sub:Ground}
In this Section we explain how a fractional topological insulator may be constructed from a set of coupled wires, as a result of a combination of spin-orbit coupling and electron-electron interaction. We start with the case of non-interacting electrons, in which case a 2D topological insulator is formed, and then introduce interactions that lead to the fractionalized phase.

\subsubsection{The integer case - a non-interacting quantum spin Hall state}
\begin{figure*}
\subfloat[\label{fig:spectrum_no_tunneling}]{\includegraphics[scale=0.15]{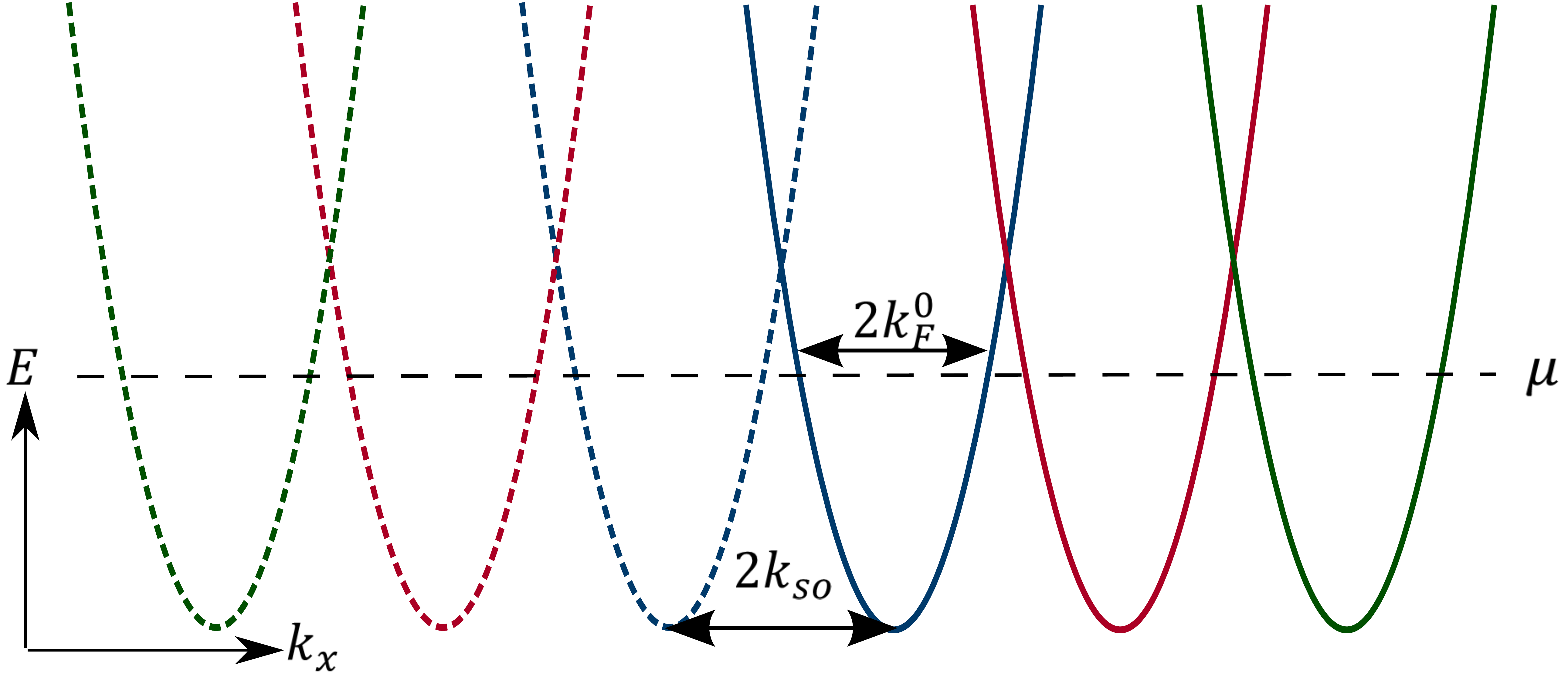}}
\subfloat[\label{fig:spectrum_no_tunneling-1}]{\includegraphics[scale=0.15]{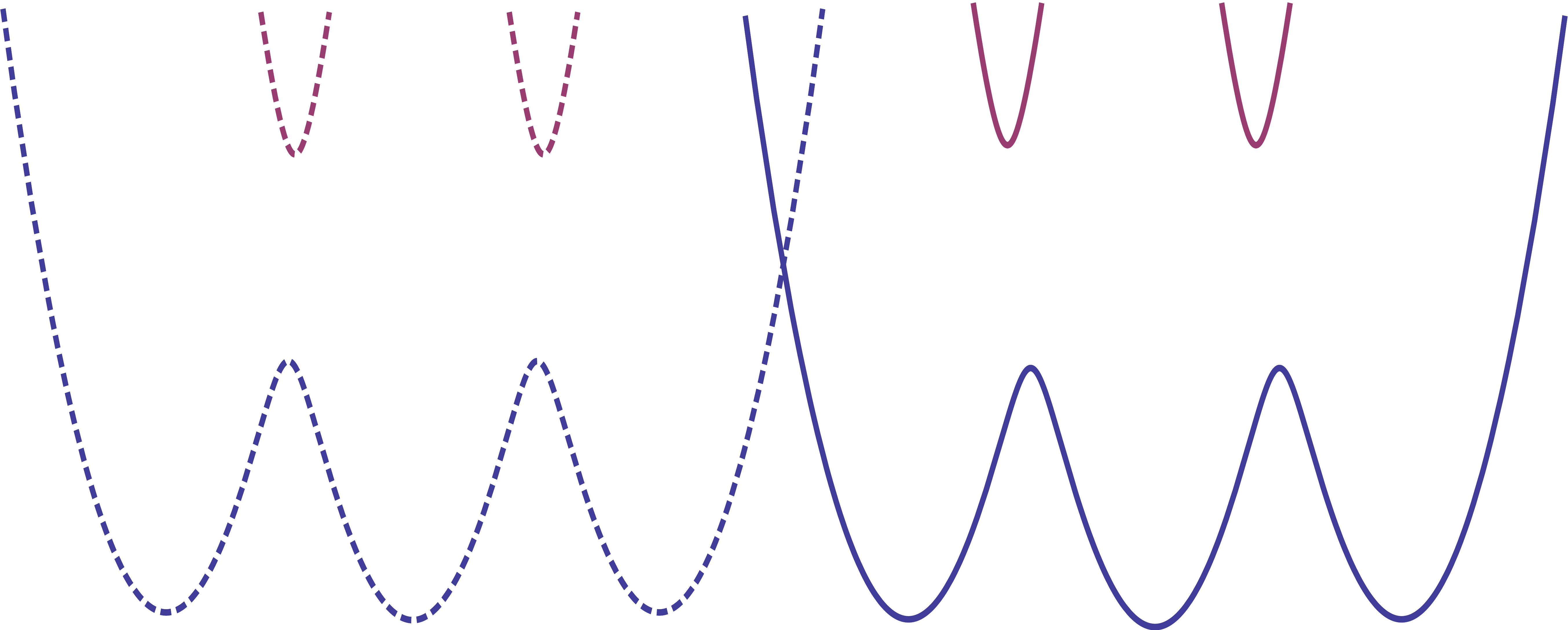}}
\protect\protect\caption{(a) The spectrum of a system consisting of three wires (see Fig. (\ref{fig:system})) with non-interacting electrons subjected to spin orbit coupling whose magnitude depends on the wire index according to Eq. (\ref{eq:HSO}),  when tunneling between
the wires is switched off. The spectra in blue, red, and green correspond
to wires number 1,2, and 3. Solid lines correspond to spin-down, and
dashed lines correspond to spin-up. (b) The resulting spectrum when
a weak spin-conserving tunneling amplitude is switched on between
the wires. The bulk is now gapped, with helical modes localized
on the edges.}
\end{figure*}
We consider an array of $N$ quantum wires, with a Rashba spin-orbit
coupling (see Fig. (\ref{fig:system})). Each wire is of length $L_x$
and has periodic boundary conditions. We tune the Rashba electric
field (which we set to be in the $y$ direction, for simplicity) such
that the spin-orbit coupling of wire number $n$ is linear with
$n$. The resulting term in the Hamiltonian takes the form
\begin{equation}
H_{so,n}=(2n-1)uk_{x}\sigma_{z},\label{eq:HSO}
\end{equation}
where $\sigma_{z}$ is the spin in the $z$ direction, and $u$ is
the spin-orbit coupling. The spectrum of wire number $n$ is therefore
\begin{equation}
E_{n}(k)=\frac{\left(k_{x}+(2n-1)k_{\rm so}\sigma_{z}\right)^{2}}{2m},\label{eq:spectrum}
\end{equation}
where $m$ is the effective mass, and $k_{\rm so}=\frac{u}{m}$. The energy
of the different wires as a function of $k_{x}$ is shown in Fig.
(\ref{fig:spectrum_no_tunneling}).

The similarity of the spectrum to the starting point of the wires
construction of the QHE \cite{Sondhi2001,Kane2002,Teo2011} is evident.
This system is then analogous to two annuli of electrons of opposite
spins subjected to opposite magnetic fields
or to the electron-hole double-layer we discussed above (see Fig. (\ref{fig:annulus_uniform})).

Following the analogy with the wires construction of the QHE, we define
the filling factor as
\begin{equation}
\nu=\frac{k_{F}^{0}}{ k_{\rm so}},\label{eq:filling fraction}
\end{equation}
where $k_{F}^{0}$ is the Fermi momentum without a spin-orbit coupling
(see Fig. (\ref{fig:spectrum_no_tunneling})).

In the ``integer'' case, $\nu=1$, the
chemical potential is tuned to the crossing points of two adjacent
parabolas.

We linearize the spectrum
around the Fermi points, and use the usual bosonization technique
to define two chiral bosonic fields $\phi_{n,\sigma}^{R/L},$ where
$n$ is the wire index, $\sigma$ is the spin index, and $R\text{ }(L)$
represents right (left) movers. In terms of these bosonic fields,
the fermion operators take the form
\begin{equation}
\psi_{n,\sigma}^{R/L}\propto e^{i\left(\phi_{n,\sigma}^{R/L}+k_{n,\sigma}^{R/L}x\right)},\label{eq:fermi operator}
\end{equation}
where
$$k_{n,\sigma}^{\rho}=-\sigma ((2n-1)k_{\rm so}+\rho k_{F}^0)$$
is the appropriate Fermi-momenta in the
absence of interactions and tunneling between the wires, with $\sigma=1$ $(-1)$ corresponding to spin up (down), and $\rho=1$ $(-1)$ corresponding to right (left) movers. The chiral
fields satisfy the commutation relations

\begin{widetext}
\begin{equation}
\left[\phi_{n\rho}^{\sigma}(x),\phi_{n'\rho'}^{\sigma'}(x')\right]=i \rho\pi\delta_{\sigma,\sigma'}\delta_{\rho,\rho'}\delta_{n,n'}\text{sign}(x-x')+i\pi  \text{sign}(n-n')+\delta_{n,n'}\pi\left(\sigma_{y}^{\sigma,\sigma'}+\delta_{\sigma,\sigma'}\sigma_{y}^{\rho,\rho'}\right).\label{eq:commutation of phi}
\end{equation}
\end{widetext}
 Eq. (\ref{eq:commutation of phi}) guarantees that the fermion fields defined in Eq. (\ref{eq:fermi operator}) satisfy Fermi-statistics.

\begin{figure*}
\includegraphics[scale=0.4]{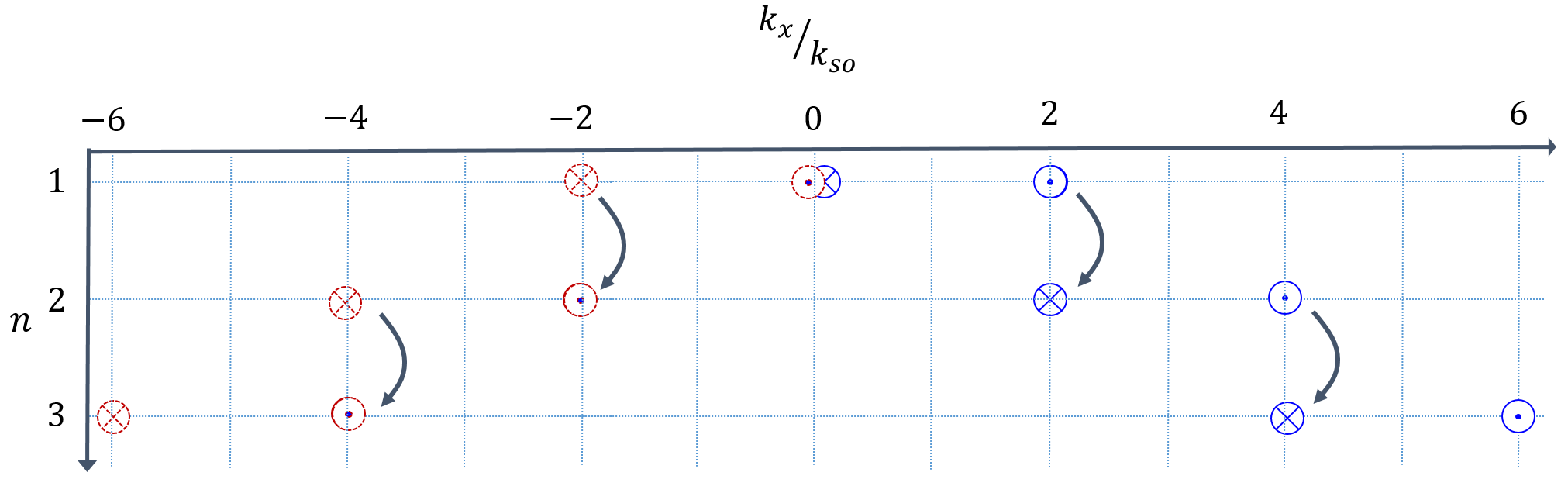}\protect\protect\caption{\label{fig:diagram-integer}A diagrammatic representation of the spectrum
in the case $\nu=1$. Once we linearize the spectrum around the Fermi-points,
it becomes convenient to plot only the Fermi-momenta as a function
of the wire index ($n$). The symbol $\odot$ ($\otimes$\textbf{)}
represents a right (left) mover. Blue (red) symbols represent the
spin-down (spin-up) component. One can observe that single electron
spin-conserving tunneling operators conserve momentum, and can therefore
easily gap out the bulk in this case. }
\end{figure*}

Once we linearize the spectrum, it becomes convenient to present it
diagrammatically by plotting only the Fermi-momenta as a function
of the wire index. Fig. (\ref{fig:diagram-integer}) shows the diagram
corresponding to $\nu=1$ , where a right (left) mover is represented
by the symbol $\odot$ ($\otimes$\textbf{)}.

One sees that single electron tunneling operators of the type
\begin{align}
H_{t\downarrow}=t\sum_{n=1}^{N-1} \int dx \left( \psi_{n+1,\downarrow}^{L\dagger}\psi_{n,\downarrow}^{R}+h.c. \right )& = \nonumber \\ t \frac{ k_{F}^{0}}{\pi} \sum_{n=1}^{N-1} \int dx \cos\left(\phi_{n+1,\downarrow}^{L}-\phi_{n,\downarrow}^{R}\right),\nonumber \\
H_{t\uparrow}= t\sum_{n=1}^{N-1} \int dx \left ( \psi_{n+1,\uparrow}^{R\dagger}\psi_{n,\uparrow}^{L}+h.c. \right ) & = \nonumber \\
 t\frac{ k_{F}^{0}}{\pi}\sum_{n=1}^{N-1} \int dx  \cos\left(\phi_{n+1,\uparrow}^{R}-\phi_{n,\uparrow}^{L}\right),\label{eq:tunneling}
\end{align}
are allowed by momentum conservation (these operators are represented by the arrows in Fig. (\ref{fig:diagram-integer})). In the above equation, we fixed the gauge for each wire such that the inter-wire tunneling takes a $\cos$ form.  Noting that these operators
commute with one another, the fields within the cosines may be pinned, and therefore the bulk is gapped. These terms, however, leave 4 gapless modes on wires 1 and
$N$: $\phi_{1,\uparrow}^{R},\phi_{1,\downarrow}^{L},\phi_{N,\uparrow}^{L},\phi_{N,\downarrow}^{R}$.
In fact, the above model is a topological insulator, and the gapless helical modes are
the corresponding edge modes, protected by time-reversal symmetry and charge conservation. Although our model also has a conservation of $S_z$, this is not actually necessary to preserve the gapless edge modes.
To completely gap out the spectrum, we have to gap out the two edges
separately. This can be done using two mechanisms: proximity coupling
of wire $1$ and $N$ to a superconductor which breaks charge conservation, or to a magnet which breaks time-reversal symmetry. The terms
in the Hamiltonian that correspond to these cases are

\begin{align}
H_{1}^{S} & =\Delta_{1} \int dx \cos\left(\phi_{1,\uparrow}^{R}+\phi_{1,\downarrow}^{L}+\delta_1\right),\nonumber \\
H_{1}^{F} & =B_{1} \int dx \cos\left(\phi_{1,\uparrow}^{R}-\phi_{1,\downarrow}^{L}+\beta_1\right),\nonumber \\
H_{N}^{S} & =\Delta_{N} \int dx \cos\left(\phi_{N,\uparrow}^{L}+\phi_{N,\downarrow}^{R}+\delta_N\right),\nonumber \\
H_{N}^{F} & =B_{N}\int dx \cos\left(\phi_{N,\uparrow}^{L}-\phi_{N,\downarrow}^{R}+\beta_N+4k_{so}Nx\right).\label{eq:gapping the edges}
\end{align}
The phases $\delta_1,\delta_N$ are the phases of the superconducting order parameter of the superconductors that couple to the wires $1,N$ respectively. The phases $\beta_1,\beta_N$ are the angles of the Zeeman fields (which lie in the $x-y$ plane) coupling to the wires $1,N$ respectively, with respect to the $x$-axis. As the last equation shows, for the magnetic field coupled to the $n$'th wire to allow for a momentum-conserving back-scattering, we must have $\beta_N=-k_{so}Nx$, i.e., the Zeeman field acting on the $N$'th wire must rotate in the $x-y$ plane at a period of $2\pi/(k_{so}N)$. This field then breaks translational invariance.

\subsubsection{The fractional case - a Fractional Topological Insulator}

\begin{figure*}
\includegraphics[scale=0.4]{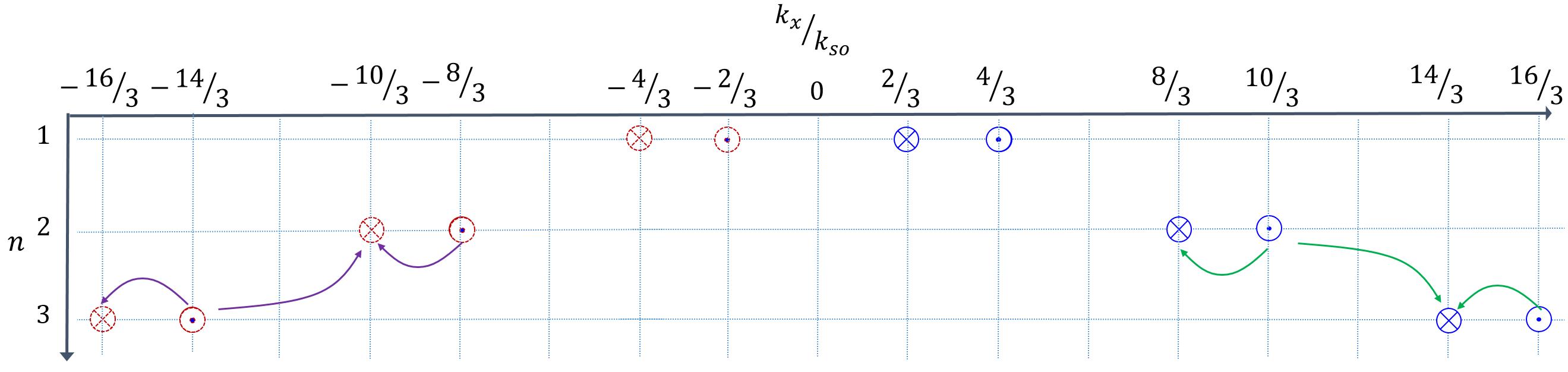}\protect\protect\caption{\label{fig:diagram-fractional}A diagrammatic representations of the
fractional case $\nu=1/3$. Now, we find that only multi-electron
processes can gap out the bulk. The processes we consider are represented
by colored arrows. In terms of the composite $\tilde{\psi}$-fields,
however, the diagram corresponding the fractional case is identical
to the one corresponding to the integer case $\nu=1$ (Fig. (\ref{fig:diagram-integer})).
In this case, the complicated multi-electron processes are transformed
into single-$\tilde{\psi}$ tunneling operators. The transformation
from $\psi$ to $\tilde{\psi}$ therefore proves very useful in analyzing
the fractional case. }
\end{figure*}

We now consider  the case $\nu=1/3$,
depicted diagrammatically in Fig. (\ref{fig:diagram-fractional}). Single electron tunneling processes of the type we considered above do not
conserve momentum (see Fig. (\ref{fig:diagram-fractional})) for this filling factor, and one
has to consider multi-electron processes in order to gap out
the bulk. The problem is simplified if one defines new chiral fermion
fields in each wire according to the transformation
\begin{equation}
\tilde{\psi}_{n,\sigma}^{R/L}=\left(\psi_{n,\sigma}^{R/L}\right)^{2}\left(\psi_{n,\sigma}^{L/R}\right)^{\dagger}\propto e^{i\left(p_{n,\sigma}^{R/L}x+\eta_{n,\sigma}^{R/L}\right)},\label{eq:composite fermions}
\end{equation}
with
\begin{align}
\eta_{n,\sigma}^{R/L} & =2\phi_{n,\sigma}^{R/L}-\phi_{n,\sigma}^{L/R},\nonumber \\
p_{n,\sigma}^{R/L} & =2k_{n,\sigma}^{R/L}-k_{n,\sigma}^{L/R}.\label{eq:etas}
\end{align}
Strictly speaking, the operators in (\ref{eq:composite fermions}) should operate at separated yet close points in space, due to the fermionic nature of $\psi_{n,\sigma}^{R/L}$.

It is simple to check that
\begin{widetext}
\begin{equation}
\left[\eta_{n\rho}^{\sigma}(x),\eta_{n'\rho'}^{\sigma'}(x')\right]=3i\rho\pi\delta_{\sigma,\sigma'}\delta_{\rho,\rho'}\delta_{n,n'}\text{sign}(x-x')+i\pi \text{sign}(n-n')+\delta_{n,n'}\pi\left(\sigma_{y}^{\sigma,\sigma'}+3\delta_{\sigma,\sigma'}\sigma_{y}^{\rho,\rho'}\right)
.\label{eq:commutation of etas}
\end{equation}
\end{widetext}

Eq. (\ref{eq:commutation of etas}) implies that $\tilde{\psi}$ satisfies Fermi statistics. In addition,
if one draws the diagram that corresponds to the $p$'s, the effective Fermi-momenta of the $\tilde{\psi}$
fields, one gets the same diagram as in the $\nu=1$ case (Fig. (\ref{fig:diagram-integer})).
The linear transformation defined in Eq. (\ref{eq:etas}) can therefore
be interpreted as a mapping from $\nu=\frac{1}{3}$ for the electrons,
to $\nu=1$ for the fermions $\tilde{\psi}$. The mapping from $\nu=1/3$ to $\nu=1$ suggests a relation between the local transformation defined in Eq. (\ref{eq:etas}) and the Chern-Simons transformation that attaches two flux quanta to each electron, making it a composite fermion. This relation will be explored in a future work \cite{Sagi2014b}. Single-$\tilde{\psi}$
tunneling operators conserve momentum, and one can repeat the process
that led to a gapped spectrum in the integer case. First, we switch
on single-$\tilde{\psi}$ tunneling operators of the form
\begin{align}
\tilde{H}_{t\downarrow}=\tilde{t}\sum_{n=1}^{N-1} \int dx \left(\tilde{\psi}_{n+1,\downarrow}^{L\dagger}\tilde{\psi}_{n,\downarrow}^{R}+h.c.\right) & = \nonumber \\ \frac{\tilde{t}}{4} \left(\frac{k_{F}^{0}}{\pi}\right)^{3} \sum_{n=1}^{N-1}\int dx \cos\left(\eta_{n+1,\downarrow}^{L}-\eta_{n,\downarrow}^{R}\right),\nonumber \\
\tilde{H}_{t\uparrow}=\tilde{t}\sum_{n=1}^{N-1}\int dx \left(\tilde{\psi}_{n+1,\uparrow}^{R\dagger}\tilde{\psi}_{n,\uparrow}^{L}+h.c.\right) & = \nonumber \\ \frac{\tilde{t}}{4}\left(\frac{k_{F}^{0}}{\pi}\right)^{3} \sum_{n=1}^{N-1} \int dx \cos\left(\eta_{n+1,\uparrow}^{R}-\eta_{n,\downarrow}^{L}\right).\label{eq:eta-tunneling}
\end{align}
While these operators are simple tunneling operators in terms of the
$\tilde{\psi}$-fields, they represent the multi-electron processes
described by the arrows in Fig. (\ref{fig:diagram-fractional}). In terms of the $\tilde{\psi}$-fields, it is clear that one cannot write analogous interactions between electrons of opposite spins, and therefore the dominating terms are those that couple electrons with the same spins. Notice
that as opposed to the integer case, these operators are irrelevant
in the weak coupling limit.
However, they may be made relevant if one
introduces strong repulsive interactions \cite{Kane2002,Teo2011,Oreg2013}, or a sufficiently strong $\tilde{t}$.

For $N$ wires, Eqs. (\ref{eq:eta-tunneling}) introduces $2N-2$ tunneling terms, which gap out $4N-4$ modes, and leave 4 gapless chiral $\eta$-modes on
the edges. Two counter-propagating modes are at the $j=1$ wire, and two are at the $j=N$ wire. Notice that the gapless $\eta$-fields on the edges are related to the corresponding $\chi$-fields defined in Sec. \ref{sec:edge modes only} by $\chi=\eta /3$. Once again, these may be gapped by proximity coupling to
a superconductor or a magnet. Operators of the type shown in
Eq. (\ref{eq:gapping the edges}), however, do not commute with the
operators defined in Eq. (\ref{eq:eta-tunneling}). The arguments of the cosines in (\ref{eq:gapping the edges}) cannot
then be pinned by Eq. (\ref{eq:eta-tunneling}). The lowest order terms that commute
with the operators in Eq. (\ref{eq:eta-tunneling}) are
\begin{align}
\tilde{H}_{1}^{S} & =\tilde{\Delta}_{1} \int dx \cos\left(\eta_{1,\uparrow}^{R}+\eta_{1,\downarrow}^{L}+\tilde{\delta_1}\right),\nonumber \\
\tilde{H}_{1}^{F} & =\tilde{B}_{1} \int dx \cos\left(\eta_{1,\uparrow}^{R}-\eta_{1,\downarrow}^{L}+\tilde{\beta_1}\right),\nonumber \\
\tilde{H}_{N}^{S} & =\tilde{\Delta}_{N} \int dx \cos\left(\eta_{N,\uparrow}^{L}+\eta_{N,\downarrow}^{R}+\tilde\delta_N\right),\nonumber \\
\tilde{H}_{N}^{F} & =\tilde{B}_{N} \int dx \cos\left(\eta_{N,\uparrow}^{L}-\eta_{N,\downarrow}^{R}+\tilde{\beta}_N+4k_{so}Nx\right).\label{eq:gapping the edges-eta}
\end{align}
Again, for the magnetic coupling to gap the edge modes on the $n$th wire, it must wind in the $x-y$ plane with a period of $2\pi/(k_{\rm so}N)$. The electronic density is three times smaller than in the previous case, so on average there is $1/3$ of an electron per period.
Guided by the analogy between the above construction and the $\nu=1/3$
FQH state on a torus, we expect the ground state to have a 3-fold
degeneracy.

Using the present formalism, will be able to see how
this degeneracy is lifted as one goes from an infinite array to the
limiting case of a few wires.

\subsubsection{Ground state degeneracy in the wire construction}

\label{sub:degeneracies in the wires construction}

For simplicity, we focus first on the FF case, where the analogy to
the FQHE on a torus is explicit. In this case, we define the idealized
Hamiltonian as
\begin{equation}
H_{I}=\tilde{H}_K+\tilde{H}_{t\uparrow}+\tilde{H}_{t\downarrow}+\tilde{H}_{1}^{F}+\tilde{H}_{N}^{F},\label{eq:idealized Hamiltonian}
\end{equation}
where
\begin{equation}
\tilde{H}_K =\frac{1}{2}\sum_{n}\sum_{\rho,\rho'}\sum_{\sigma,\sigma'}\int dx\left(\partial_{x}\eta_{n\rho}^{\sigma}\right)V_{\rho,\rho'}^{\sigma,\sigma'}\left(\partial_{x}\eta_{n\rho'}^{\sigma'}\right) \label{eq:kinetic term}
\end{equation}
is the quadratic term that contains the non-interacting part of the Hamiltonian, and small momentum interactions (for simplicity, we consider only intra-wire small momentum interactions).
We assume that all the inter-wire terms become relevant and acquire an expectation
value. To investigate the properties of the ground state manifold,
we define the two unitary operators
\begin{widetext}
\begin{align}
U_{y}(x) & =e^{i\frac{1}{3}\left(\sum_{n=1}^{N}\left(\eta_{n,\uparrow}^{R}-\eta_{n,\uparrow}^{L}+\eta_{n,\downarrow}^{R}-\eta_{n,\downarrow}^{L}\right)\right)}= e^{i\upsilon(x)}e^{i\frac{1}{3}\left(\eta_{N,\downarrow}^{R}-\eta_{N,\uparrow}^{L}+\eta_{1,\uparrow}^{R}-\eta_{1,\downarrow}^{L}\right)},\label{eq:Uy} \\
U_{x} & =e^{i\frac{1}{3}\int_{0}^{L}\partial_{x}\eta_{1,\uparrow}^{R}dx}.\label{eq:Ux}
\end{align}
\end{widetext}
All the $\eta$ fields are functions of position $x$.
The phase $\upsilon(x)$ in Eq. (\ref{eq:Uy}) is given by
\[\]
\begin{equation}
\upsilon(x)=\frac{1}{3}\left[ \sum_{n=1}^{N-1}\left(\eta_{n+1,\uparrow}^{R}-\eta_{n,\uparrow}^{L}\right)-\sum_{n=1}^{N-1}\left(\eta_{n+1,\downarrow}^{L}-\eta_{n,\downarrow}^{R}\right)\right].\label{eq:upsilon}
\end{equation}
Since all the operators in the sum are pinned by the bulk Hamiltonian, they may be treated as classical fields, and their value becomes $x$-independent in any one of the ground states. Similarly, the combination of operators $(\eta^R_{N, \downarrow} - \eta^L_{N, \uparrow} + \eta^R_{1, \uparrow} -\eta^L_{1,\downarrow})$ which appears on the right side of Eq. (\ref{eq:Uy}) is pinned by the coupling to the boundary, and becomes independent of $x$. Therefore, the operators $U_y(x)$ may be considered to be independent of $x$ within the manifold of ground states.

Notice that the second equality in Eq. (\ref{eq:Uy}) shows that $U_y(x)$ defined in terms of the wires degrees of freedom is identical to Eq. (\ref{eq:Uy in terms of chi}) (up to a phase). The
form of $U_{y}(x)$ shown in the first equality of Eq. (\ref{eq:Uy}) is useful because it allows us to express $U_{y}(x)$ as a product of electronic operators:
\begin{equation}
U_{y}=e^{i\left(\sum_{n=1}^{N}\left(\phi_{n,\uparrow}^{R}-\phi_{n,\uparrow}^{L}+\phi_{n,\downarrow}^{R}-\phi_{n,\downarrow}^{L}\right)\right).}\label{eq:O in terms of phi}
\end{equation}
where the $x$-dependence of the operators is omitted for brevity.
It can be verified that
\begin{equation}
[U_y(x),U_y(x')]=0
\label{Uyxxprime}
\end{equation}
and that
\begin{equation}
\left[U_{x},H_{I}\right]=\left[U_{y},H_{I}\right]=0,\label{eq:o and u with H}
\end{equation}
so that operating $U_{y}(x)\text{ or }U_{x}$ on a ground state leaves the system  in the ground state manifold. Using Eq. (\ref{eq:commutation of etas}),
it can also be checked directly that
\begin{equation}
U_{x}U_{y}(x)=U_{y}(x)U_{x}e^{\frac{2\pi}{3}i}.\label{eq:commuting o and u}
\end{equation}
independent of $x$.
The smallest  representation of this algebra requires $3\times3$ matrices
\cite{Nayak2008}, which shows that the ground state of the idealized Hamiltonian (\ref{eq:idealized Hamiltonian}) must be at least
3-fold degenerate.

The operators $U_{y}$ ($U_{x}$) can be interpreted
as the creation of a quasiparticle-quasihole pair, tunneling of the
quasiparticle across the $y$ ($x$) direction of the torus and annihilating
the pair at the end of the process. %
In fact, if we adopt this interpretation, Eq. (\ref{eq:commuting o and u})
is a direct consequence of the fractional statistics of the quasiparticles
\cite{Nayak2008}.

A similar analysis can be carried out for the SS case. $U_{x}$ is
identical to the operator used in the FF case, but now $U_{y}$ takes
the form
\begin{equation}
U_{y}=e^{i\frac{1}{3}\left(\sum_{n=1}^{N}\left(\eta_{n,\uparrow}^{R}-\eta_{n,\uparrow}^{L}+\eta_{n,\downarrow}^{L}-\eta_{n,\downarrow}^{R}\right)\right),}\label{eq:Uy in the SS case}
\end{equation}
and the entire analysis can be repeated.

\subsection{The coupled wires construction of an electron-hole double layer}

\begin{figure*}
\subfloat[\label{fig:specrum_electron-hole}]{\includegraphics[scale=0.15]{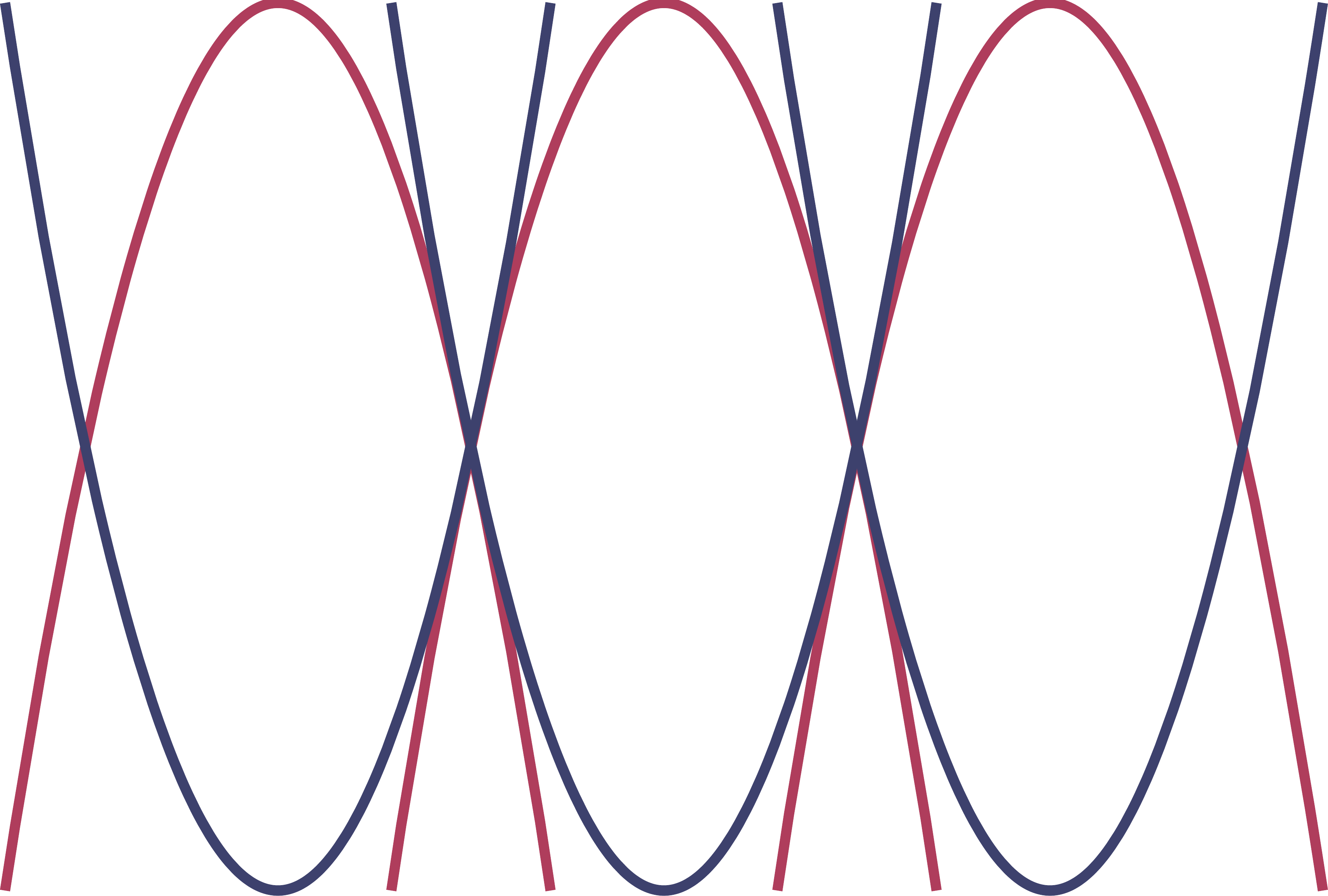}

}\subfloat[\label{fig:spectrum_electron-hold-gapped}]{\includegraphics[scale=0.15]{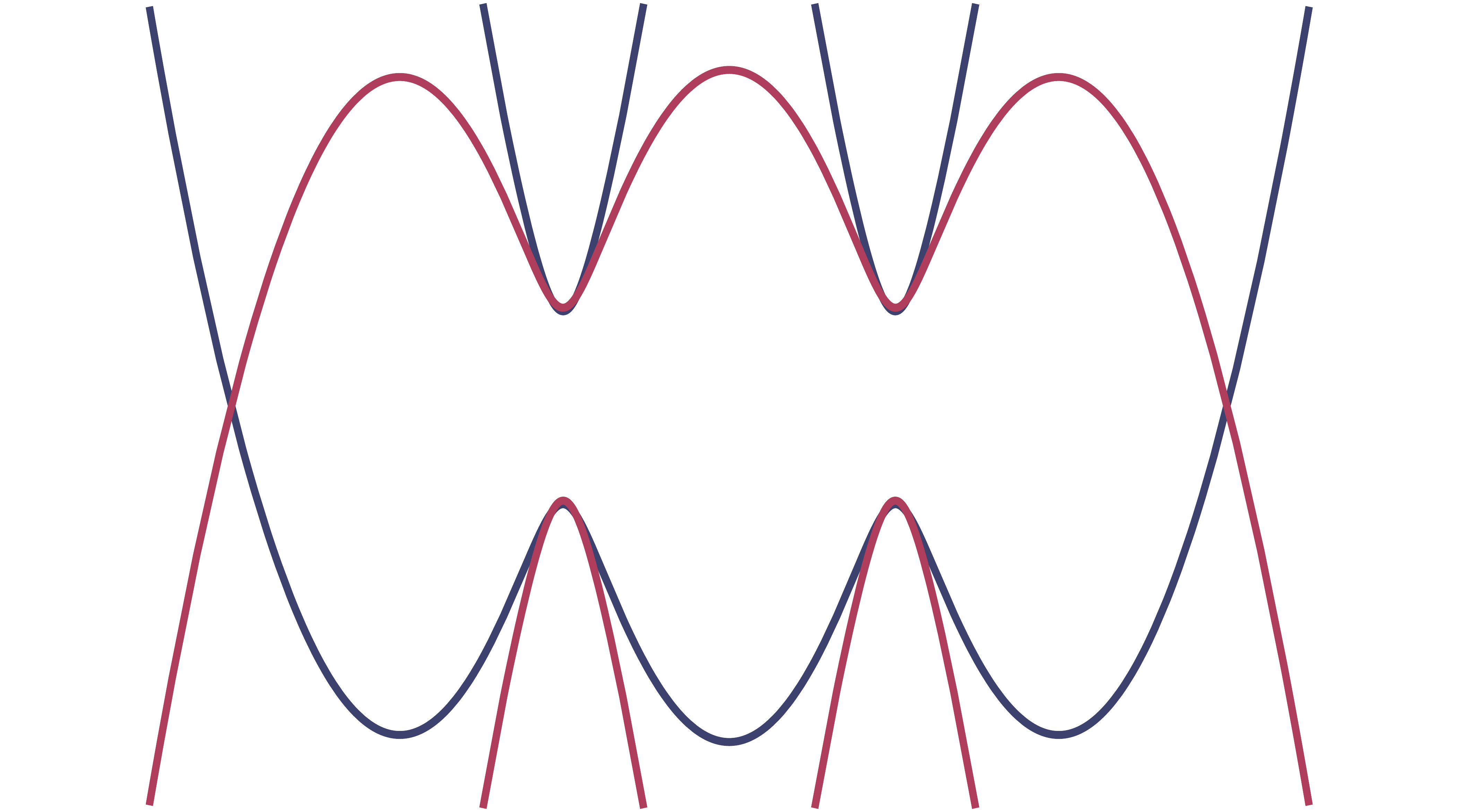}

}\subfloat[\label{fig:spectrum_electron-hole-fractional}]{\includegraphics[scale=0.3]{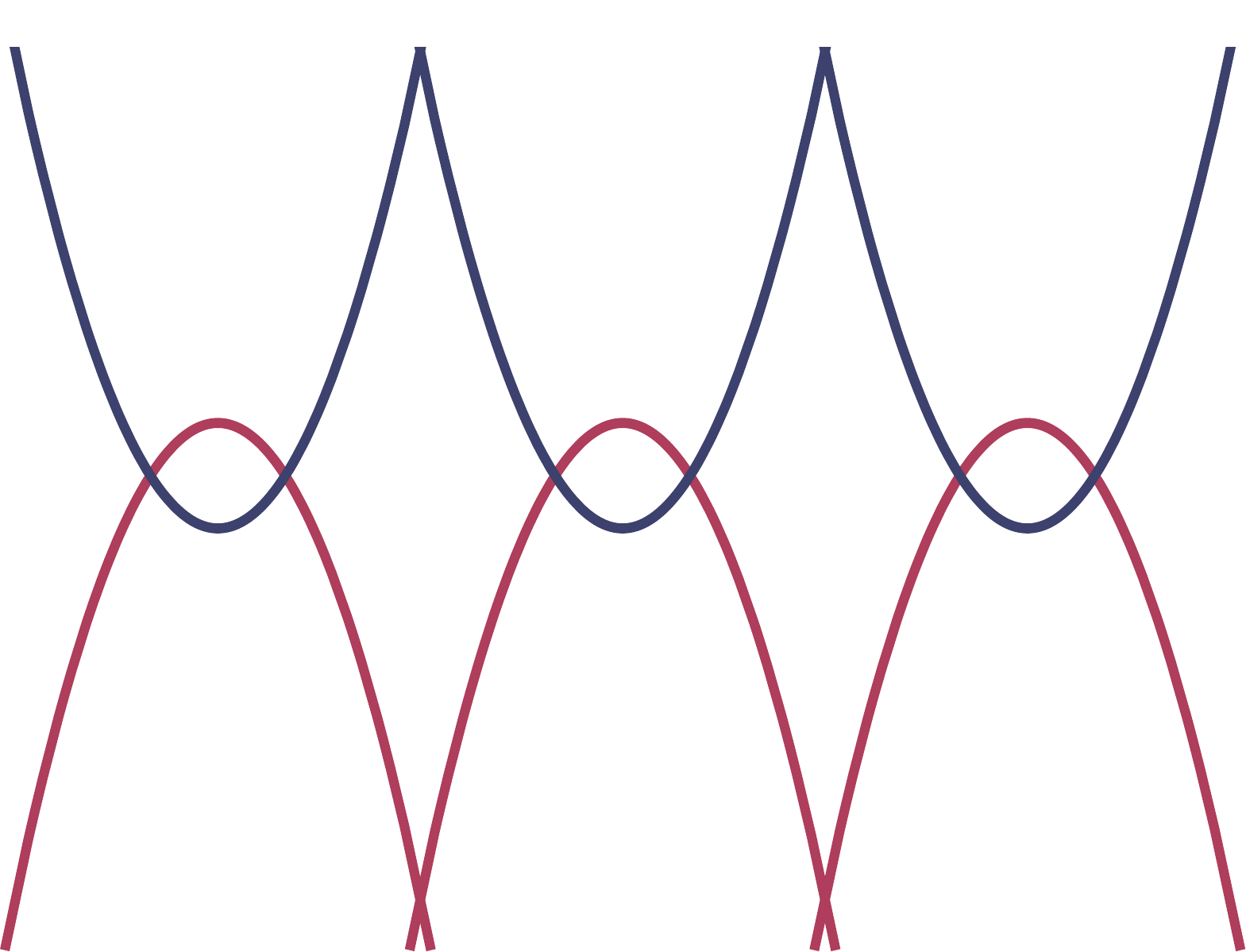}

}\caption{(a) The spectrum of the wires model for an electron-hole double layer
at filling $\nu=1$ when all the inter-wire terms are switched off.
The spectra in blue correspond to wires in the electron layer, and
the spectra in red correspond wires in the hole layer. (b) The spectrum
when tunneling between wires in the same layer is switched on. A gap
is formed in the bulk, and we get achiral edge modes. (c) The spectrum
in the fractional case $\nu=1/3$.}
\end{figure*}

In this Section we explain how one can model a quantum Hall
electron-hole double layer at a fractional filling factor $\nu=1/3$ using a set of coupled wires. Most of the
analysis is very similar to the analysis presented for the fractional
topological insulator, but some technical differences are worth pointing
out.

We examine a system with two layers, each containing an array of wires.
In one layer, the electron layer, we tune
the system such that only states near the bottom of the electronic
band are filled. In this case, we can approximate the spectra of the
various wires as parabolas. If we add a constant magnetic field $B$
perpendicular to the layers, and use the Landau gauge to write the electromagnetic potential as $\mathbf{A}= -B y \hat{x}$, the entire band structure of wire number
$n$ will be shifted by an amount $2k_{\phi}n$, where $k_{\phi}$
is defined as $k_{\phi}=\frac{eBa}{2\hbar}$. The energy of wire number
$n$ is therefore written in the form (if we choose the position of wire number 1 to be at $y=a/2$)
\begin{equation}
E_{n}(k)=\frac{\left(k_{x}-\right(2n-1\left)k_{\phi}\right)^{2}}{2m}+U_{e},\label{eq:spectrum of electron wire}
\end{equation}
where $U_{e}$ is a constant term, and $m$ is the effective mass.
In the hole layer the bands of the various wires are nearly
filled, such that we can expand the energy near the maximum as
\begin{equation}
E_{n}(k)=-\frac{\left(k_{x}-\right(2n-1\left)k_{\phi}\right)^{2}}{2m}+U_{h}.\label{eq:spectrum of hole wire}
\end{equation}
In the above, we assumed that the effective masses of the
electron and the hole layers have the same magnitude and opposite
signs. We assume that $U_{h}>U_{e},$ and tune the chemical potential
to be $\mu=\frac{U_{e}+U_{h}}{2}$. Defining $\delta\epsilon=\frac{U_{h}-U_{e}}{2}$, we get the spectra
\begin{align}
E_{n}(k)-\mu=\left[\frac{\left(k_{x}-\right(2n-1\left)k_{\phi}\right)^{2}}{2m}-\delta\epsilon\right]\sigma, \label{eq:electron hole specturm}
\end{align}
where $\sigma=1(-1)$ for the electron (hole) layer. This way the
system has a built-in particle-hole symmetry in its low energy Hamiltonian.
Notice that as a result of the magnetic field, the spectra of the
two layers are shifted in the same direction. This is a consequence of the common origin of the electron and hole spectra from a
Bloch band whose shift is determined by the direction of the magnetic
field.

We define $k_{F}^{0}=\sqrt{2m\delta\epsilon}$, and the filling factor
is now given by $\nu=\frac{k_{F}^{0}}{k_{\phi}}$. In the case $\nu=1$,
the corresponding spectrum is given by Fig. (\ref{fig:specrum_electron-hole}).
As before, if we apply tunneling between adjacent wires in the same
layer, we get the gapped spectrum in Fig. (\ref{fig:spectrum_electron-hold-gapped}).
Furthermore, we see that each edge carries a pair of counter propagating
edge modes (one for each layer).

It is straightforward to generalize this to the case of filling $\nu=1/3$,
shown in Fig. (\ref{fig:spectrum_electron-hole-fractional}). To treat
this case, we follow exactly the same steps as in Sec. \ref{sub:Ground}:
we first linearize the spectrum, and write the problem in terms of
the chiral bosonic degrees of freedom $\phi_{n,\sigma}^{R/L}$, where
now $\sigma=e,h$ represents the layer number, and $n$ represents
the wire index. To treat the fractional case, we define new chiral
fields $\eta_{n,\sigma}^{R/L}=2\phi_{n,\sigma}^{R/L}-\phi_{n,\sigma}^{L/R}$.
Like before, it can be checked that these modes behave like modes
at filling 1, so we can repeat the analysis performed in this case.

This process leaves us with two counter propagating $\eta$-modes
on each edge: $\eta_{1,e}^{L},\eta_{1,h}^{R},\eta_{N,e}^{R},\eta_{N,h}^{L}$.
These modes can be gapped out by terms analogous to the terms in Eq.
(\ref{eq:gapping the edges-eta}):
\begin{align}
\tilde{H}_{1}^{S} &=\tilde{\Delta}_{1}\cos\left(\eta_{1,e}^{L}+\eta_{1,h}^{R}+\tilde{\delta}_{1}\right), \nonumber \\
\tilde{H}_{1}^{F} &=\tilde{B}_{1}\cos\left(\eta_{1,e}^{L}-\eta_{1,h}^{R}+\tilde{\beta}_{1}\right), \nonumber \\
\tilde{H}_{N}^{S} &=\tilde{\Delta}_{N}\cos\left(\eta_{N,e}^{R}+\eta_{N,h}^{L}+\tilde{\delta}_{N} +4k_{\phi}Nx\right),\nonumber \\
\tilde{H}_{N}^{F} &=\tilde{B}_{N}\cos\left(\eta_{N,e}^{R}-\eta_{N,h}^{L}+\tilde{\beta}_{N}\right).\label{eq:gapping the edges electron hole}
\end{align}
In contrast to the case of the fractional topological insulator, here
the backscattering terms conserve momentum, i.e., do not include phases that are linear in $x$. Rather, the superconducting
term $\tilde{H}_{N}^{S}$ appears not to conserve momentum. However, the flux between the two superconductors will lead to a winding
of the phase difference between them, which can cancel the $x$-dependent
phase of $H_N^S$.

Let us first consider the situation where the bounding superconductor
wires are thin enough that there are no vortices inside them.   The energy
of a superconducting ring is minimized when $\Delta \phi$, the change in
the superconducting phase around the ring is equal to $2 e \Phi $, where
$\Phi$ is the magnetic flux enclosed by a circle embedded at the center of
the wire.  The value of  $\Delta \phi$ is quantized in multiples of $2
\pi$, and in practice there may exist a number of metastable states where
it differs from $2 e \Phi $ by a finite amount and the wire carries a
supercurrent around its circumference.  Let us consider a model where
there is a distance $a'$ between the center of the inner most
superconductor and the center of our first electron-hole nanowire and a
similar separation between the $N$th nanowire and the outer
superconductor.  If the centers of the nanowires are separated from each
other by a distance $a$, then the flux $\Phi$ is equal to $Ba L_x (N-1 + 2
(a'/a))$.  In this case, if the superconductors are in their ground
states, we get $\tilde{\delta}_{1}= \left(-2+4 \frac{a'}{a}\right)k_{\phi}x+\tilde{\delta}_{1}^0$ and $\tilde{\delta}_{N}= -\left(4N-2+4\frac{a'} {a}\right)k_{\phi}x+\tilde{\delta}_{N}^0$, where $\tilde{\delta}_{1(N)}^0$ do not depend on $x$. If $a'$ is tuned to $a'=a/2$, the oscillating phases are eliminated from Eq. (\ref{eq:gapping the edges electron hole}).

If $a'$ differs from $a/2$, it may be still possible to gap out the edges.
 If the phase mismatch is small, and if coupling to the superconductor is
not too weak, then there can be an adjustment of the electron and hole
occupations in the nanowires nearest the two edges, which allows the phase
change around the nanowires to match the phase change in the
superconductors. The energy gain due to formation of a gap can exceed
the energy cost of altering the charge densities in the nanowires.

If the difference between $a'$ and $a/2$ is too large, then carrier
densities in the inner and outer nanowires will not change enough to
satisfy the phase matching condition.  In this case, a variation of the magnetic field of order $1/N$ would eliminate the $x$-dependence of the phases at the cost of introducing quantum Hall quasiparticles in the bulk of the system. For large $N$, the density of these quasiparticles will be small. Presumably they will become localized and not take the system out of the quantum Hall plateau.

We note that the separation $a'$ can be engineered, and, in principle can
even be made negative. Consider, for example, a situation where the superconducting wire sits above the plane of the nanowires, so that depending on the shape of a cross-section of the wire, its center of gravity may sit inside or outside of the line of contact to the outermost nanowire.}

The situation is more complicated if the superconductors are thick enough
that they contain vortices in the presence of the applied magnetic field.
If the vortices are effectively pinned, however, it should be possible to
achieve conditions where the electron-hole system is gapped and
experiments such as Josephson current measurements can be performed.

The degeneracy of the ground states in both the SS and FF cases may be shown by defining the two operators $U_{x}$ and
$U_{y}$ in exactly in the same form as we did in Sec. \ref{sub:degeneracies in the wires construction}
(with $\downarrow\rightarrow e$ and $\uparrow\rightarrow h$), and
following the same analysis.

\section{Measurable imprint of the topological degeneracy in quasi-one dimensional systems}

\label{sec:crossover}

We now look at the quantum Hall double-layer system with $\nu=\pm 1/3$. As long as the bulk gap does not close, in the limit of infinite $L_x$ and infinite $N$ (or $L_y$) we expect deviations from the idealized Hamiltonian not to couple the three ground states. When $N$ and $L_y$ are finite and $L_x$ is still infinite, coupling does occur, and the degeneracy is lifted. 

Generally, hermitian matrices operating within the $3\times 3$ subspace of ground states of the idealized Hamiltonian may all be written as combination of nine unitary matrices $O_{j,k}$
\begin{equation}
\Delta H=\sum_{(j,k)=(-1,-1)}^{(1,1)}\lambda_{j,k}O_{j,k} .\label{eq:general hermitian}
\end{equation}
where
\begin{equation}
O_{jk}=U_{x}^{j}U_{y}^{k}.\label{eq:general unitary}
\end{equation}
and $\lambda_{jk}=\lambda_{-j,-k}^*e^{-\frac{i2\pi jk}{3}}$. 
Note that a direct consequence of Eq. (\ref{eq:commuting o and u}) is that $U_{x}^{3}=U_{y}^{3}=1$
(this can most easily be understood by recalling that the operators
transport quasiparticles across the torus. Acting three times with
each of them is equivalent to transporting an electron around the
torus, which cannot take us from one ground state to another). However, in the limit of infinite $L_x$ local operators cannot distinguish between states of different fractional charges, and therefore cannot contain the operator $U_x$.  Thus, up to an unimportant constant originating from $\lambda_{00}$, deviations from the idealized Hamiltonian (projected to the ground state manifold) take the form of Eq. (\ref{deviation}):

\begin{equation}
\Delta H= \lambda U_{y}+\lambda^* U_{y}^{\dagger},\label{eq:term in the Hamiltonian due to Uy}
\end{equation}
The coefficient $\lambda$ may be expressed as an integral,
\begin{equation}
\lambda = \int
dx T(x),
\label{localtunnelingintegrated}
\end{equation}
and  we expect that the  absolute value of the amplitude $T(x)$
should fall off exponentially with $N$, as discussed in Section II B.
One can see this explicitly in the various models we have constructed from
coupled wires.  For example, in the case of a fractional topological
insulator with magnetic boundaries, the operator $U_y$, according to (32 )
and (17)  involves a product of factors involving four electronic creation
and annihilation operators on each of the $N$ wires.  The bare Hamiltonian
contains only four-fermion operators on a single wire, and two-fermion
operators that connect adjacent wires, with an amplitude $t$ that we
consider to be small.  The operator $U_y$ can only be generated by higher
orders of perturbation theory, in which the microscopic tunneling
amplitude $t$ occurs at least $2N$ times.  In our analysis, we have assumed
that interaction strengths on a single wire are comparable to the Fermi
energy $E_F$, so we expect $T$ to be of order   $|t/E_F|^{2N}$ or smaller.
Similar arguments apply to the other cases of superconducting boundaries
or electron-hole wires.  We also note that if the system is time-reversal
invariant, we must have $T=T^*$.

The phase of $T(x)$  depends on the realization - electron-hole quantum Hall {\it vs.} fractional topological insulator - and on the gapping mechanism - two superconductors or two magnets. We start from the case of the fractional topological insulator gapped by two superconductors. Eqs. (\ref{eq:gapping the edges-eta}) shows that for the edges to be gapped, the superconductors on the two edges should have uniform phases $\tilde{\delta}_1,\tilde{\delta}_N$. We choose a gauge where $\tilde{\delta}_1=0$ and denote $\varphi=\tilde{\delta}_N$ to be the phase difference.

In the case of a fractional topological insulator, the proximity gapping terms are
\begin{align}
\tilde{H}_{1}^{S} & =\tilde{\Delta}_{1}\int dx \cos\left(\eta_{1,\uparrow}^{L}+\eta_{1,\downarrow}^{R}+\varphi\right),\nonumber \\
\tilde{H}_{N}^{S} & =\tilde{\Delta}_{N}\int dx \cos\left(\eta_{N,\uparrow}^{R}+\eta_{N,\downarrow}^{L}\right)\label{eq:Hs with relative phase}
\end{align}
(note that these terms involve coupling to the superconductor, and we therefore have $\tilde{\Delta}_{1(N)} \propto |\Delta_{1(N)}|$, where $\Delta_{1(N)}$ are the corresponding superconducting order parameters). 
{We define new bosonic fields through the additional transformation
\begin{equation}
\tilde{\eta}_{1,\uparrow}^{L}=\eta{}_{1,\uparrow}^{L}+\frac{\varphi}{2},\ \ \tilde{\eta}{}_{1,\downarrow}^{R}=\eta{}_{1,\downarrow}^{R}+\frac{\varphi}{2},\label{eq:transformed etas}
\end{equation}
and $\tilde{\eta}{}_{n,\sigma}^{\rho}=\eta{}_{n,\sigma}^{\rho}$ for all the other values of $n,\sigma,\rho$. If we rewrite the Hamiltonian in terms of the new fields, the phase $\varphi$ is  eliminated
from the idealized Hamiltonian. However, this modifies the operator
$U_{y}$ (defined in Eq.(\ref{eq:Uy})), which now takes the form
\begin{equation}
U_{y}=e^{i\frac{1}{3}\left(\sum_{n=1}^{N}\left(\tilde{\eta}_{n,\uparrow}^{R}-\tilde{\eta}_{n,\uparrow}^{L}+\tilde{\eta}_{n,\downarrow}^{L}-\tilde{\eta}_{n,\downarrow}^{R}\right)\right)}e^{i\frac{\varphi}{3}}.\label{eq:Uy after gauge transformation}
\end{equation}
Thus, a non-zero phase difference $\varphi$ shifts the argument of $\lambda$ in Eq. (\ref{eq:term in the Hamiltonian due to Uy}) by $\frac{\varphi}{3}$. In the time reversal symmetric case $\lambda$ is real, and we find, by diagonalizing $\Delta H$, that
\begin{align}
\Delta E_{1} & =2\lambda L_{x}\cos\left(\frac{\varphi}{3}\right),\nonumber \\
\Delta E_{2} & =2\lambda L_{x}\cos\left(\frac{\varphi-2\pi}{3}\right),\nonumber \\
\Delta E_{3} & =2\lambda L_{x}\cos\left(\frac{\varphi+2\pi}{3}\right).\label{eq:spectrum as a function of phi}
\end{align}

The resulting spectrum as a function of $\varphi$ is depicted in
Fig. (\ref{fig:spectrum with flux}).

At $\varphi= \pi n$ the
degeneracy is not completely lifted, as two states remain 2-fold
degenerate. These states are not coupled by the low energy Hamiltonian (\ref{eq:term in the Hamiltonian due to Uy}) and the lifting of their degeneracy requires terms of $j\ne 0$ in (\ref{eq:general unitary}). Such terms distinguish between states of different edge charges $f_i$ and originate from tunneling between the three physically distinct minima of the potential (\ref{eq:cosine perturbation}).
The amplitude for tunneling, and hence the splitting, is proportional to $e^{-S}$, with $S$ the imaginary action corresponding to the tunneling trajectory. Due to the integration over $x$ in the Hamiltonian, this action is linear in $L_x$, and hence the tunneling amplitude scales as $e^{-(L_x/\xi_x)}$.
Neglecting this splitting, Eq. (\ref{eq:spectrum as a function of phi})
shows that all eigenstates have a $6\pi$ periodicity. A measurement of the Josephson current, given by the derivative of the energy with respect to $\varphi$, can detect the $6\pi$-periodicity. Due
to the exponentially small splitting at the crossing points, this
property can be observed by changing the flux at a rate that is not slow enough to follow this splitting.

Note that the $6\pi$-periodic component of the spectrum is completely determined by Eq.
(\ref{eq:commuting o and u}). This part of the spectrum is therefore
highly insensitive to the microscopic details, and can serve as a
directly measurable imprint of the topological degeneracy with only
a few wires. There will also be a contribution from ordinary Cooper pair tunneling between the superconductors, which does not distinguish between the ground states and has $2 \pi$ periodicity. This term will alter the detailed shapes of the three spectra but not their splitting or periodicity. In the case where time reversal symmetry does not hold, $\lambda$ is not necessarily real. Consequently, the spectrum in Eq. (\ref{eq:spectrum as a function of phi}) is shifted according to  $\varphi \rightarrow \varphi+\text{Arg}\left(\lambda\right)$, and the crossing points are not constrained to be at $\varphi=\pi n$.

{Similar results arise in the FF case for a quantum Hall electron-hole double layer. Now, the angle $\varphi$
is the relative orientation angle of the Zeeman fields (which lies
in the $x-y$ plane). To be precise, if we fix the Zeeman field at wire number $N$ to point at the $x$
direction, and the field at wire number 1 to have an angle $\varphi$
relative to the $x$ direction, we get the proximity terms
\begin{eqnarray}
\tilde{H}_{1}^{F} & = & \tilde{\lambda}_{1F}\int dx\cos\left(\eta_{1,\uparrow}^{L}-\eta_{1,\downarrow}^{R}+\varphi\right),\nonumber \\
\tilde{H}_{N}^{F} & = & \tilde{\lambda}_{NF}\int dx\cos\left(\eta_{N,\uparrow}^{R}-\eta_{N,\downarrow}^{L}\right).\label{eq:FF case proximity terms with phi}
\end{eqnarray}
Similar to Eq. (\ref{eq:transformed etas}), we define new bosonic fields through the
transformation
\begin{equation}
\tilde{\eta}_{1,\uparrow}^{L}=\eta{}_{1,\uparrow}^{L}+\frac{\varphi}{2},\ \ \tilde{\eta}{}_{1,\downarrow}^{R}=\eta{}_{1,\downarrow}^{R}-\frac{\varphi}{2},\label{eq:transformed etas-1}
\end{equation}
and $\tilde{\eta}{}_{n,\sigma}^{\rho}=\eta{}_{n,\sigma}^{\rho}$ for
the other fields. Again, the gapping term acting on the $N$'th wire returns to its original
form (with $\varphi=0$), but $U_{y}$ becomes $U_{y}e^{i\frac{\varphi}{3}}$. Therefore, the spectrum as a function of $\varphi$ is identical to the spectrum found in the SS case.}

In the other two cases, the situation is more complicated, since $\varphi$ depends on $x$. For the quantum Hall electron-hole double layer gapped by superconductors $\varphi$  increases linearly with $x$, due to the flux penetrating the junction between the two superconductors. For the fractional topological insulator gapped by magnets, Eq. (\ref{eq:gapping the edges-eta}) requires that $\beta_N$ increases linearly with $x$. In both cases, this winding leads to $\lambda=\int dx |t(x)|e^{i 2\pi n x/L+i\varphi}$, with $n$ an integer. A uniform tunneling amplitude $|t(x)|$ then leads to a vanishing $\lambda$, while non-uniformity allows for a non-vanishing $\lambda$.

\section{Extensions to other Abelian states}

\label{sec:extension}

We have shown above that it is possible effectively realize experimentally
the $\nu=\frac{1}{3}$ FQHE state on a torus, and that by measurement
of the Josephson effect in the resulting construction we can directly
measure the corresponding topological degeneracy. In this section we extend the above results to other Abelian FQHE
states.

For a FQHE state described by a $M\times M$ {\it K}-matrix, there is a ground
state degeneracy of $d=\det K$ on a torus, and
$d$ topologically distinct quasiparticles. Each quasiparticle is a multiple of the minimally charged quasiparticle, whose
charge is $e^{*}=\frac{e}{d}$.

Repeating the analysis we carried out in Sec. \ref{sec:model}, we consider an electron-hole double layer system or a fractional topological insulator, and couple the counter-propagating edge modes. Since there are now $M$ pairs of counter-propagating modes on each edge, we need $m$ scattering terms. We assume that these terms are all mutually commuting, that they are either all charge-conserving or all superconducting, and that the $M$ edge modes of each layer (spin-direction) are mutually coupled. Under these assumptions, each of the four edges is characterized by a single quantum number - the fractional part of the total charge $f_i$ (with $i=1,\cdots,4$), which may take the values $-\frac{d-1}{2d},-\frac{d-3}{2d},\cdots,\frac{d-1}{2d}$. The mutual coupling between the $M$ edge modes excludes the possibility of other quantum numbers being constants of motion.   Similar to the case where $\nu = 1/3$, the requirements of a total integer charge for each layer or spin direction, together with the mechanism of gapping and the requirement to minimize the energy of the edge Hamiltonians, relate all values of $f_i$ to one another.

We work in a basis $\left|f\right\rangle $ where the fractional charges
$f_i$ are well defined. We define the unitary operator $U_{y}$ which
transfers a single minimally charged quasiparticle, analogously to
the operator defined before, such that $U_{y}\left|f\right\rangle =\left|\left(f+e^*/e\right)\rm{mod}(1)\right\rangle $.
It follows that $U_{y}^{l}\left|f\right\rangle =\left|\left(f+l e^*/e\right)\rm{mod}(1)\right\rangle $,
and that $U_{y}^{d}=1$. We therefore have in general
\begin{equation}
\left(U_{y}^{l}\right)^{\dagger}=U_{y}^{d-l}.\label{eq:Um dagger}
\end{equation}

Again, in the quasi-1D limit where $L_x$ is infinite and $N$ is finite, Hermitian combinations of the
operators $U_{y}^{l}$ are the only operators capable of lifting the degeneracy. The amplitude of these terms
falls exponentially with $N$.
In order to analyze the effects of these perturbations we consider
terms of the form
\begin{equation}
\Delta H=\sum_{l=1}^{(d-1)/2} \left(\lambda_{l} U_{y}^{l}e^{i\delta_{l}}+h.c.\right),\label{eq:delta_H-1}
\end{equation}
where $\lambda_{l}\propto e^{-N/\xi_{l}}$ is a real coefficient (note that we expect terms with $l>1$ to result from higher orders in $e^{-N}$. More specifically, we expect $\xi_l \propto \frac{1}{l}$). The summation was terminated at $(d-1)/2$ because of Eq.
(\ref{eq:Um dagger}) and the requirement that the Hamiltonian is hermitian. Again, the resulting spectrum depends on the realization, the gapping mechanism, and the uniformity of the tunneling amplitude. This dependence is similar to the one discussed for $\nu=1/3$. For example, for uniform tunneling between two superconductors separated by a fractional topological insulator, a relative phase $\varphi$ between the two
superconductors translates to $\delta_l=\varphi\frac{e^{\star}}{e}l$.

The spectrum of this Hamiltonian for the time reversal symmetric case is then
\begin{equation}
\Delta E_{p}=2\sum_{l=1}^{(d-1)/2}\lambda_{l}\cos\left(\frac{l}{d}\left(\varphi+2 p \pi\right)\right),\label{eq:spectrum K matrix}
\end{equation}
with $p=1\ldots d$. Each eigenstate has a $2\pi d$-periodicity, and like the $\nu=1/3$ case
we find that the overall periodicity is $2\pi$ times the degeneracy
of the system in the thermodynamic limit. In
addition, similar to the $\nu=1/3$ case, at the time-reversal invariant
points $\varphi=\pi n$, we have degeneracy points protected
by the length of the wires. For example, at $\varphi=0$, we have
$\frac{d-1}{2}$ pairs of states $\left|p\right\rangle ,$$\left|d-p\right\rangle $ $(p=1,\ldots \frac{d-1}{2}$)
which have the same energy. It can easily be checked from Eq. (\ref{eq:spectrum K matrix})
that the same number of crossings occurs for any $\varphi=\pi n$. Hence if the spectrum is measured, the degeneracy $d$ can found by simply counting the number of crossing points at $\varphi=\pi n$. Note that due to the terms with $l>1$, we can have additional crossing points at $\varphi \neq n \pi$. Again, if time reversal symmetry does not hold the crossing points can be shifted. One can still show that in the most general case there must be at least the same number of crossing points as the number of crossing points at $\varphi = \pi n$ in the time reversal invariant case. The smallest number of degeneracy points occurs when the functions $\Delta E_{p}$ have a single maximum and a single minimum between 0 and $2\pi d$. In that case, the energies that correspond to two different values of $p$ must cross at two points between 0 and $2\pi d$. We therefore have 2 crossing points for each pair $p_1,p_2$, The total number of degeneracy points, summed over all the pairs $p_1,p_2$ is therefore  $2\left(\begin{array}{c}
d\\
2
\end{array}\right)=d(d-1),$
which is the number of crossing points at all the values $\varphi=\pi n$ in the time reversal invariant case.
Depending on the values of $\lambda_l$, we may have more than a single minimum and a single maximum, in which case we can get additional crossing points.

As an example we examine the case $\nu=2/5$, which can be characterized
by the $K$-matrix
\begin{equation}
K=\left(\begin{array}{cc}
3 & 2\\
2 & 3
\end{array}\right).\label{eq:K matrix}
\end{equation}

The degeneracy on a torus in this case is $d=5$ and the spectrum (in the time reversal invariant case)
is
\begin{equation}
\Delta E_{p}=2\lambda_{1}\cos\left(\frac{1}{5}\left(\varphi+2 p\pi\right)\right)+2\lambda_{2}\cos\left(\frac{2}{5}\left(\varphi+2p\pi\right)\right),\label{eq:spectrum K matrix-1}
\end{equation}
with $p=1\ldots 5$. If we take for example $\lambda_{2}/\lambda_{1}=0.2$, the resulting spectrum
is shown in Fig. (\ref{fig:spectrum nu=2/5}).
\begin{figure}
\includegraphics[scale=0.27]{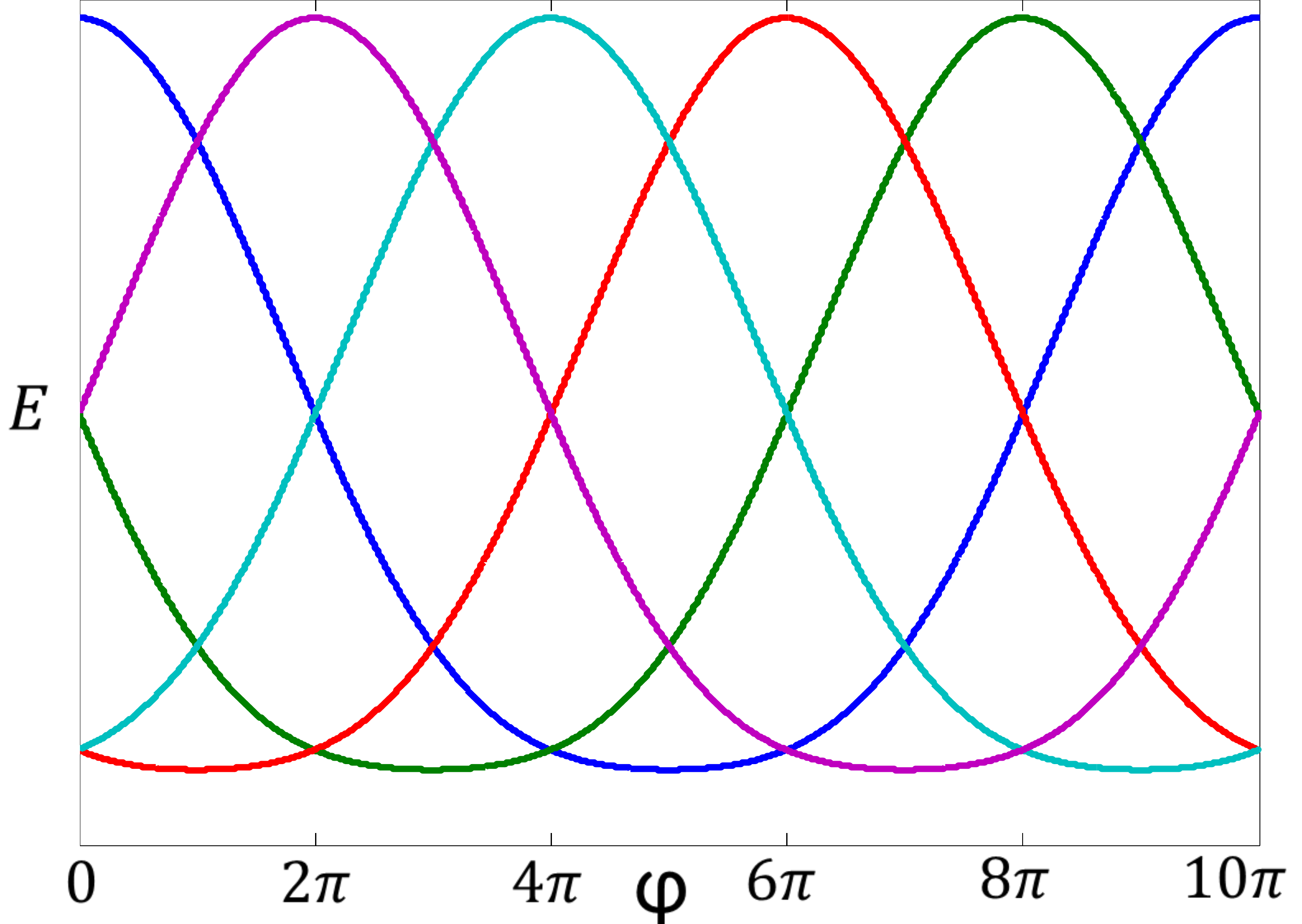}\protect\caption{\label{fig:spectrum nu=2/5} The spectrum corresponding to $\nu=2/5$ with $\lambda_2/\lambda_1=0.2$ as a function of the relative phase difference $\varphi$.  The periodicity of each eigenstate is $10 \pi$. At the points $\varphi=\pi n$, we find two crossing points whose splitting falls exponentially with $L_x$.}
\end{figure}

\section{Conclusions}
\label{sec:conclusions}
The topological degeneracy on a torus is perhaps the defining property of a fractionalized phase, and the most prominent signature of a topological order. As such, it is unfortunate that for the most accessible fractionalized phase - the Fractional Quantum Hall effect - it is impossible to to directly create a toroidal geometry, that requires magnetic monopoles. In this work we study two annular geometries that are topologically equivalent to that of a torus. One geometry is based on an electron-hole double layer where the electrons and the holes are at fractional quantum Hall states of opposite filling fractions. The other is based on a fractional topological insulator at which the two spin directions of the electrons are at fractional quantum Hall states of opposite filling fractions. Both geometries carry counter-propagating edge modes on the interior and the exterior edges of the annuli, and these edge modes may be coupled and gapped in two mechanisms - back-scattering and proximity coupling to superconductors.

Considering the two dimensional regime where the annuli are too wide to have a significant coupling between the interior and the exterior edges, we established here the topological degeneracy that characterizes each of the geometries we consider, and their dependence on the gapping mechanism on each of the edges. Furthermore, we used the quantum number of the fractional charge or dipole on each of the edges to characterize the ground states.

In the regime where the annuli are narrow such that the interior and the exterior are coupled, the degenerate ground states split in energy. Searching for remnants of the topological order that survive the transition to the quasi-one dimensional regime, we studied the dependence of the spectrum of split ground states as a function of the phase difference between the two superconductors or the relative angle between the direction of magnetization of the two magnets. We find that the spectrum includes points in which the splitting is exponentially small in the circumference of the annulus, and thus is not split when the width becomes small.

At finite temperature there will be thermally excited pairs of quasiparticles and quasiholes in the bulk. When reaching the edge, these excitations carry the potential of introducing transitions between the states that cross at Figs. (\ref{fig:spectrum with flux}) and (\ref{fig:spectrum nu=2/5}). The density of these quasiparticles and the resulting transition rates are expected to be exponentially small at low temperatures.

The spectra of Figs. (\ref{fig:spectrum with flux}) and (\ref{fig:spectrum nu=2/5}) give rise to a remarkable experimental consequence. As long as experiments are done on timescales at which the exponentially small transitions between states at the crossing points may be neglected, the Josephson effects give a $2\pi d$-periodicity, where $d$ is degeneracy in the 2D thermodynamic limit. Despite the fact that the degeneracy was lifted in the quasi-1D regime, it leaves an imprint in the Josephson effect. 


\begin{acknowledgments}
We are indebted to Erez Berg and Eran Sela for insightful and important conversations. This work was supported by Microsoft's Station Q, the US-Israel Binational Science Foundation, the Israel Science Foundation (ISF), the
Minerva Foundation, and the European
Research Council under the European Community's
Seventh Framework Program (FP7/2007-2013)/ERC
Grant agreement No. 340210.

 \end{acknowledgments}

\begin{appendix}
\section{Projection of Hermitian matrices onto the ground state manifold}
In section \ref{sec:crossover} we stated that hermitian matrices operating within the $3 \times 3$ subspace of ground states of the idealized Hamiltonian may all be written as a combination of nine unitary matrices as in Eq.~(\ref{eq:general hermitian}). In this appendix we prove this statement. To do so, we show that any operator acting on this subspace can be written as a combination of $O_{j,k}$. It follows that in particular any hermitian operator can be written in the form shown in Eq. (\ref{eq:general hermitian}), if the constraint $\lambda_{jk}=\lambda_{-j,-k}^*e^{-\frac{i2\pi jk}{3}}$ is imposed.

The operator $U_x$ defined in Eq. (\ref{eq:Ux}) measures the charge on the edge modes. We expect that it will have three eigenvalues corresponding to edge charges equal to $0, 1/3$, and $-1/3$.
We denote the eigenstate with zero charge on the edge by $\left| 0 \right\rangle$. In addition we introduce the notation: $\left| 1 \right\rangle =U_y \left| 0 \right\rangle$ and $\left| -1 \right\rangle =U_y \left| 1\right\rangle = U_y^2 \left| 0 \right\rangle$.

By definition $U_x \left| 0 \right\rangle = \left| 0 \right\rangle $ and it is easy to check, using the identity $U_x U_y = e^{i \alpha} U_y U_x,\; \alpha=2\pi/3$, that:
\begin{equation}
U_x \left| 1 \right\rangle=U_x U_y \left| 0 \right\rangle =e^{i\alpha} U_y U_x \left| 0 \right\rangle
= e^{i \alpha} U_y \left| 0 \right\rangle  = e^{i \alpha} \left| 1 \right\rangle
\end{equation}
and similarly we find
\begin{equation}
U_x \left| -1 \right\rangle = e^{2 i \alpha} \left| -1 \right\rangle = e^{-i \alpha} \left| -1 \right\rangle.
\end{equation}
The set $\left| -1 \right\rangle, \left| 0 \right\rangle, \left| 1 \right\rangle$ forms a complete basis for the $3 \times 3$ subspace of ground states so that any operator $\hat O$, projected onto this subspace, can be written in this basis as:
\begin{equation}
\label{eq:sum}
\hat O = \sum_{j,l= -1,0,1} \left| j \right\rangle \left\langle j \right| O \left| l \right\rangle   \left\langle l \right|.
\end{equation}
Since
\begin{equation}
U_x= e^{-i \alpha}\left| -1 \right\rangle \left\langle -1 \right|+ \left| 0 \right\rangle \left\langle 0 \right| + e^{+i \alpha}\left| 1 \right\rangle \left\langle 1 \right|
\end{equation}
we find that (notice that $\cos \alpha= \cos 2 \alpha \text{ for } \alpha = 2\pi/3$ )
\begin{equation}
\label{eq:00}
\left| 0 \right\rangle \left\langle 0 \right| = (U_x+U_x^\dagger-2\cos \alpha \mathds{1})/(2(1-\cos \alpha))
\end{equation}
with $\mathds{1}= \left| -1 \right\rangle \left\langle -1 \right|+ \left| 0 \right\rangle \left\langle 0 \right| + \left| 1 \right\rangle \left\langle 1 \right|$ being a unit matrix in the $3 \times 3$ subspace. All the other  $\left| j \right\rangle \left\langle l \right|$  operators in the expansion of Eq.~(\ref{eq:sum}) can be obtained by multiplying  the presentation of $\left| 0 \right\rangle \left\langle 0 \right|$ in Eq.~(\ref{eq:00}) by $U_y$ or $U^\dagger_y$ from left or right. For example:
\begin{equation}
\left| 1 \right\rangle \left\langle 0 \right| = U_y \left| 0 \right\rangle \left\langle 0 \right|=
U_y U_x+U_y U_x^\dagger-2\cos \alpha U_y.
\end{equation}
Hence the expansion of Eq.~(\ref{eq:general hermitian}) follows.

\end{appendix}
 \bibliographystyle{apsrev4-1}
%

\end{document}